\documentclass{ws-ijgmmp}

\usepackage{graphicx}  
\usepackage{latexsym}
\usepackage{amssymb}
\usepackage{amsmath}

\def\beq{\begin{equation}}
\def\eeq{\end{equation}}
\def\rmd{{\rm d}}
\def\three{{}^{(3)}\kern-1.5pt } 
\def\four{{}^{(4)}\kern-1.5pt }
\def\h{h}
\begin{document}

\markboth{D. Bini, E. Bittencourt, A. Geralico and R.T. Jantzen}
{Slicing black hole spacetimes}

%%%%%%%%%%%%%%%%%%%%% Publisher's Area please ignore %%%%%%%%%%%%%%%
%
\catchline{}{}{}{}{}
%
%%%%%%%%%%%%%%%%%%%%%%%%%%%%%%%%%%%%%%%%%%%%%%%%%%%%%%%%%%%%%%%%%%%%

\title{SLICING BLACK HOLE SPACETIMES
}

\author{\footnotesize DONATO BINI}

\address{
Istituto per le Applicazioni del Calcolo ``M. Picone,'' CNR, I--00185 Rome, Italy\\
ICRA, ``Sapienza'' University of Rome,  I--00185 Rome, Italy\\
INFN - Sezione di Napoli, Complesso Universitario di Monte S. Angelo, Via Cintia Edificio 6, 80126 Napoli, Italy\\
\email{binid@icra.it}
}

\author{\footnotesize EDUARDO BITTENCOURT}

\address{
CAPES Foundation, Ministry of Education of Brazil, Bras\'ilia, Brazil\\
ICRA, ``Sapienza'' University of Rome,  I--00185 Rome, Italy\\
Physics Department, \lq\lq Sapienza" University of Rome, I-00185 Rome, Italy\\
\email{eduardo.bittencourt@icranet.org}
}

\author{\footnotesize ANDREA GERALICO}

\address{
Istituto per le Applicazioni del Calcolo ``M. Picone,'' CNR, I--00185 Rome, Italy\\
ICRA, ``Sapienza'' University of Rome,  I--00185 Rome, Italy\\
INFN - Sezione di Napoli, Complesso Universitario di Monte S. Angelo, Via Cintia Edificio 6, 80126 Napoli, Italy\\
\email{geralico@icra.it}
}

\author{\footnotesize ROBERT T. JANTZEN}

\address{
Department of Mathematics and Statistics, Villanova University, Villanova, PA 19085, USA\\
ICRA, ``Sapienza'' University of Rome,  I--00185 Rome, Italy\\
\email{robert.jantzen@villanova.edu}
}

\maketitle

\begin{history}
\received{(Day Month Year)}
\revised{(Day Month Year)}
\end{history}

\begin{abstract}
A general framework is developed to investigate the properties of useful choices of stationary spacelike slicings of stationary spacetimes whose congruences of timelike orthogonal trajectories are interpreted as the world lines of an associated family of observers, the kinematical properties of which in turn may be used to geometrically characterize the original slicings. On the other hand properties of the slicings themselves can directly characterize their utility motivated instead by other considerations like the initial value and evolution problems in the 3-plus-1 approach to general relativity. An attempt is made to categorize the various slicing conditions or  ``time gauges" used in the literature for the most familiar stationary spacetimes: black holes and their flat spacetime limit.
\end{abstract}


\section{Introduction}

A century after the birth of general relativity, we now take for granted the existence of various stationary spacelike slicings of stationary spacetimes which have certain special geometrical properties useful in studying the astrophysical consequences of say, black hole spacetimes. Many of these slicings arise from geometrical properties of their irrotational congruences of orthogonal timelike trajectories, interpreted as the world lines of an associated family of observers which may be either geodesic or accelerated. Other slicings are instead characterized by the intrinsic or extrinsic geometry of the slicing itself. We here survey the various categories of such useful slicings for nonrotating and rotating black hole spacetimes, but starting with the limiting flat Minkowski spacetime which allows the greatest variety of examples of special slicings.

Among the observer-defined slicings of black hole spacetimes is the Boyer-Lindquist time coordinate slicing associated with the usual stationary accelerated observers referred to equivalently as fiducial or locally non-rotating or zero angular momentum observers, abbreviated as FIDOs, LNOs or ZAMOs. Geodesic observer families instead characterize the rain, drip and hail coordinate systems \cite{TaylorWheeler,Finch} which include  the Painlev\'e-Gullstrand coordinates \cite{Painleve21,Gullstrand22} and their generalizations \cite{HL04}. For nonrotating black holes the latter are also characterized by the intrinsic curvature properties of their associated slicing, whose induced geometry is flat. Smarr and York \cite{smarr-york} pioneered linking the preservation of kinematical properties of the slicing to the choice of time lines in an evolving spacetime, while constant mean curvature slicings were seen as privileged from the point of view of the initial value problem even earlier \cite{Malec:2003dq,Malec:2009hg,Schinkel:2014dya}.
Special spherically symmetric slicings of the nonrotating Schwarzschild black hole spacetime were first considered by Estabrook et al \cite{Estabrook:1973ue,geyer}.
Recent work has investigated  the geometry associated with analogue black holes \cite{BLV05,V04} and shown that neither intrinsically flat nor conformally flat slicings of the Kerr spacetime exist \cite{Doran00,GP00,Kroon03,HL04}.

On the other hand, the analysis of the Cauchy problem of general relativity has also led to the introduction of  spacetime slicings useful in  simplifying the evolution equations, like harmonic slicing \cite{bona} (and closely related time gauges  \cite{cook}),  defined by requiring the time coordinate of the spacetime associated with the slicing to be a harmonic function.
Furthermore, the numerical relativity study of multi-black hole dynamics \cite{cook-review} takes advantage of the use of ``hyperbolic slicings," requiring spatial compactification techniques at infinity \cite{Ohme:2009gn,Zenginoglu:2010zm,Schinkel:2013zm}, as well as horizon penetrating coordinates like Painlev\'e-Gullstrand coordinates.
We refer to these as ``analytic slicings," belonging to the Cauchy problem literature, in contrast with those of a ``geometrical" nature.

Motivated primarily by a desire to better understand the underlying geometrical structure of these spacetimes, we systematically review, develop and discuss  special slicings of black hole spacetimes together with their flat Minkowski limit.
We use geometric units with $c=1=G$. Greek indices run from $0$ to $3$ and refer to spacetime quantities, while Latin indices from $1$ to $3$ and refer to spatial quantities.

\section{Spacetime, general coordinates and spacelike $3$-surfaces}

Let  $\{t,x^a\}$ be a generic set of coordinates adapted to a slicing of spacetime  by spacelike hypersurfaces of constant values of the time coordinate $t$ and write the metric as
\beq
\label{metrica2}
{}^{(4)}\rmd s^2=g_{\alpha\beta}\rmd x^\alpha \rmd x^\beta
=g_{tt} \rmd t^2+2g_{ta}\rmd t {\rmd x}^a+g_{ab}\rmd x^a \rmd x^b, \qquad (a,b=1,2,3)\,.
\eeq
The convenient Wheeler lapse-shift notation \cite{MTW} re-expresses the metric in the form 
\begin{eqnarray}
\label{metrica2bis}
{}^{(4)}\rmd s^2&=&-N^2 \rmd t^2+g_{ab}(\rmd x^a +N^a \rmd t)(\rmd x^b+N^b \rmd t)\nonumber\\
&\equiv&-N^2 \rmd t^2+g_{ab}\omega^a\omega^b\,,
\end{eqnarray}
defining the lapse function  $N$ and shift vector field $N^c$ which satisfy 
\beq
\label{lap_shi}
g_{tt}=-(N^2-N_cN^c)\,, \qquad g_{ta}=N_a=g_{ab}N^b\,,
\eeq
while the 1-forms $\{\omega^a\}$ are orthogonal to the unit timelike 1-form $n^\flat =-N\rmd t$\footnote{
The symbol $A^\flat$  denotes here the completely covariant form of a tensor $A$.
} 
associated with the unit normal vector field $n$ to the slicing
\beq
\label{nZAMO}
n=\frac{1}{N}(\partial_t -N^a\partial_a)\,.
\eeq
Here we use the equivalent shortened notations for the the spatial coordinate frame  $\partial_a \equiv \partial_{x^a}\equiv e_a$.
The induced Riemannian $3$-metric on the time slices is simply 
\beq
\label{3met}
\three\rmd s^2=g_{ab}\rmd x^a \rmd x^b\,,
\eeq
with spacetime volume element $\sqrt{-g}=N\sqrt{{\rm det}(g_{ab})}$.
Note that $\{N\,dt,\omega^a\}$ is the dual frame to the frame $\{n,\partial_a\}$ reflecting the orthogonal decomposition of each tangent space adapted to the slicing and its normal direction. 

For a stationary spacetime, one can choose these coordinates so that the stationary symmetry corresponds to translation in the time coordinate $t$, with associated Killing vector field $\partial_t$. To transform to another stationary slicing, without loss of generality one can consider restricted choices of the new spacelike time coordinate of the form
\beq\label{eq:Tt}
T= t-f(x^1,x^2,x^3)
\eeq 
which retain the time lines of the original coordinate system if the spatial coordinates are not changed. One is still free to choose new time lines by changing the spatial coordinates as well, but unless the time lines are associated with a Killing vector field, the metric will become explicitly time-dependent.

On a generic slice $\Sigma$ of the new slicing, the 1-form
\beq
\label{dz2}
\rmd T=\rmd t -f_a\rmd x^a\,, \qquad f_a=\partial_a f\,,
\eeq
vanishes identically, i.e.,
\beq
\label{dzSigma2}
\rmd T\big|_{\Sigma}=0 \quad\rightarrow\quad  \rmd t=f_a\rmd x^a \quad ({\rm on} \,\, \Sigma)\,.
\eeq
If we retain the spatial coordinates and only introduce this new time coordinate, 
\beq\label{eq:tTf}
t=T+f(X^1,X^2,X^3), \qquad x^a=X^a,
\eeq
then one must distinguish the spatial coordinate frame vector fields tangent to the old and new time coordinate hypersurfaces
\beq
\partial_T=\partial_t\,, \qquad 
E_a\equiv\partial_{X^a}=\partial_{a}+f_a\partial_t \,,\qquad
\rmd X^a = \rmd x^a
\,.
\eeq
Re-expressing the spacetime metric then leads to
\begin{eqnarray}
\label{metrica}
{}^{(4)}\rmd s^2&=&-L^2 \rmd T^2 +\h_{ab}(\rmd X^a +L^a \rmd T) (\rmd X^b +L^b \rmd T)\,, \nonumber \\
&\equiv& -L^2\rmd T^2 +\h_{ab} \Omega^a \Omega^b\,,
\end{eqnarray}
where $\h_{ab}$ (the induced metric on $\Sigma$) and $L$ and $L_c$ (the new lapse and shift) are given by
\begin{eqnarray}
\label{gammaab}
\h_{ab}&=&g_{ab}+g_{ta}f_b+g_{tb}f_a+g_{tt}f_af_b\,,  \nonumber \\
L_c&=& g_{tc}+g_{tt}f_c\,, \quad L^c=\h^{cd}L_d \,, \quad L^2= -g_{tt}+L_cL^c\,.
\end{eqnarray}
Similarly to Eq.~(\ref{lap_shi}), we then have
\beq\label{eq:gLL}
g_{TT}=-(L^2-L_cL^c)=g_{tt}, \qquad g_{Ta}=L_a=\h_{ab}L^b= g_{ta}+g_{tt}f_a\,.
\eeq
One can also evaluate the contravariant metric
\beq
g^{TT}=-\frac{1}{L^2}\,, \qquad g^{Ta}=\frac{L^a}{L^2}\,, \qquad \h ^{ab} = g^{ab}
\eeq
while the spacetime and spatial metric determinants satisfy  
\beq
\sqrt{-g}=L\sqrt{\h}\,,\qquad
\h={\rm det}(\h_{ab})\,.
\eeq

Note that $\{L\,\rmd T,\Omega^a\}$ is the dual frame to the frame $\{\mathcal{N},E_a\}$ adapted to the orthogonal slicing decomposition of the tangent space and that on $\Sigma$ one has $\Omega^a\big|_{\Sigma} = \rmd X^a =\rmd x^a$. Any ``spatial tensor" has only Latin indexed components allowed to be nonzero in this frame. 

The unit timelike 1-form normal to the $T=const$ slicing is given by
\beq
\label{calNflat}
{\mathcal N}^\flat=-L \,\rmd T=-L (\rmd t -f_a\rmd x^a)\,,
\eeq
with associated unit timelike normal vector field
\beq
{\mathcal N}=\frac{1}{L}(\partial_T-L^aE_a)\,.
\eeq
In turn ${\mathcal N}$ can be expressed in terms of $n$ as
\beq
{\mathcal N}=\gamma ({\mathcal N},n)[n+\nu({\mathcal N},n)]\,, 
\eeq
where $\nu({\mathcal N},n)$ is the relative velocity of ${\mathcal N}$ with respect to $n$ and $\gamma ({\mathcal N},n)$ the associated gamma factor, explicitly
\beq\label{eq:gammanu}
\gamma ({\mathcal N},n)=\frac{L}{N}(1+N^df_d)\,, \qquad
\nu({\mathcal N},n)^\flat=\frac{Nf_a}{(1+N^cf_c)}\omega^a\,,
\eeq
as follows from Eq.~(\ref{calNflat}) after re-expressing $\rmd t$ and $\rmd x^a$ in terms on $n^\flat$ and $\omega^a$.
A straightforward calculation shows that the new lapse function and shift vector field for the same time coordinate lines are given by
\begin{eqnarray}
L&=&\gamma({\mathcal N},n) N[1-\nu({\mathcal N},n)_cN^c]=\frac{\gamma({\mathcal N},n) N}{1+N^cf_c}\,, \nonumber\\ 
\vec L&=&L^a\partial_a=L\,\gamma({\mathcal N},n)\left(\frac{N^a}{N}-\nu({\mathcal N},n)^a\right)\partial_a\,.
\end{eqnarray}
The above decomposition gives a more transparent kinematical meaning to the various quantities, as we will show below. 

Starting from the spacetime unit volume 4-form $\eta_{\mu\alpha\beta\gamma}$, one can associate with any timelike unit vector field $U$, whether $n$ or $\cal{N}$,  a spatial volume 3-form $\eta(U)_{\alpha\beta\gamma}=U^\mu \eta_{\mu\alpha\beta\gamma}$ which can be used to define the cross product $\times_U$ and the curl operator ${\rm curl}_U$ in the local rest space $LRS_U$ of $U$, as well as a spatial duality operation for antisymmetric spatial tensor fields. In a spatial frame adapted to that subspace, for spatial vector fields $X$ and $Y$ in that subspace, one has 
\beq
  [ X \times_U Y ]^a = \eta(U)^a{}_{bc} X^b Y^c\,,\quad
 [{\rm curl}_U\, X]^a =\eta(U)^a{}_{bc} \nabla(U)^b X^c\,,
\eeq
where the spatial covariant derivative $\nabla(U)$ of any tensor field (including $U$) is obtained by projecting all indices  of the spacetime covariant derivative of that tensor into the local rest space using the associated projection tensor
$P(U)^\alpha{}_\beta = \delta^\alpha{}_\beta+U^\alpha U_\beta$ whose fully covariant form $P(U)^\flat $ is the spatial metric. 

We conclude this section by introducing the relevant tensor quantities needed to evaluate both the intrinsic and extrinsic curvature of a typical slice $\Sigma$ as well as provide a  geometrical characterization of the kinematical properties of its normal congruence ${\mathcal N}$.

\begin{enumerate}

\item Intrinsic curvature of $\Sigma$.

This is obtained evaluating the Riemann tensor components $R(\h)_{abcd}$ of the $3$-metric $\h$ induced on $\Sigma$, i.e.,
\beq
\label{3geom}
\three\rmd s^2=\h_{cd} \,\rmd X^c \rmd X^d\,.
\eeq
Note that in the three-dimensional case the Riemann tensor is completely determined by the associated Ricci tensor.

\item Conformal flatness of $\Sigma$.

The Cotton tensor associated with the induced 3-metric (\ref{3geom}) is given by
\beq
C(\h)^{a}{}_{bc}=2\left(R(\h)^{a}{}_{[b}-\frac14 R(\h)\delta^a{}_{[b}\right){}_{;c]}\,,
\eeq
where all operations including the covariant derivative $f_{|a}=\nabla(\mathcal{N})_a f$ refer to the 3-metric.
A vanishing Cotton tensor characterizes the conformal flatness of the spatial metric.
Taking the spatial dual of the two covariant indices gives the equivalent Cotton-York tensor
\beq
\label{CYtensdef}
C(\h)^{ab}=\frac12  R(\h)^{a}{}_{cd}\,\eta(\mathcal{N})^{bcd}\,,
\eeq
which is symmetric because of the twice contracted Bianchi identities of the second kind
\beq
C(\h)^{ab}=C(\h)^{(ab)}\equiv Y(\h)^{ab}\,,
\eeq
where
\beq
Y(\h)^{ab}=\frac12 C(\h)^{(a}{}_{cd}\eta(\mathcal{N})^{b)cd}=[{\rm Scurl}_{\mathcal{N}}\, R(\h)]^{ab}\,,
\eeq
and the symmetric tensor curl ``Scurl" operation \cite{mfg,cotton_bob,cotton_bob2} is defined by
\beq
[{\rm Scurl}_{\mathcal{N}}\, R(\h)]^{ab}=\eta(\mathcal{N})^{dc(a}\nabla(\mathcal{N})_d R(\h)^{b)}{}_c\,.
\eeq
The spatial metric $\h$ is conformally flat if the spatial Ricci tensor $R(\h)$ has vanishing Scurl.

\item Extrinsic curvature of $\Sigma$.

This is obtained evaluating the Lie derivative of the spacetime metric along the unit normal ${\mathcal N}$ to $\Sigma$ and projecting the result orthogonally to ${\mathcal N}$ and raising an index to make a mixed tensor, with an extra numerical factor
\beq
K({\mathcal N})^\alpha{}_\beta
 =-\frac12 P({\mathcal N})^{\alpha\gamma} \, \left[ \pounds_{\mathcal N}\, g \right]_{\gamma\delta} P(\mathcal{N})^\delta{}_\beta\,.
\eeq
Its trace ${\rm Tr}\,[K(\mathcal{N})]=K(\mathcal{N})^\beta{}_\beta$ is the mean curvature of the slice.
The constant mean curvature (CMC) time gauge is a slicing with constant mean curvature on each slice, though it may vary from slice to slice.
A maximal slicing instead has vanishing mean curvature on every slice.
When the tensor $K(\mathcal{N})$ itself vanishes, the slicing is called {\it totally geodesic}, or {\it extrinsically flat}.
An invariant characterization of the extrinsic curvature can be obtained by studying its eigenvalues, namely those of the $3\times 3$ matrix of components $(K({\mathcal N})^a{}_b)$ in an adapted frame since it is a spatial tensor.
The three eigenvalues in turn can be expressed in terms of the three scalar trace invariants of the powers of the extrinsic curvature ${\mathcal T}_1\equiv{\rm Tr}\,[K({\mathcal N})]$, ${\mathcal T}_2\equiv{\rm Tr}\,[K({\mathcal N})^2]$ and ${\mathcal T}_3\equiv{\rm Tr}\,[K({\mathcal N})^3]$.

\item Kinematical fields associated with  ${\mathcal N}$: acceleration, vorticity, expansion and shear.

These are obtained by decomposing the covariant derivative of ${\mathcal N}$ into its irreducible parts under a change of frame 
\beq
\nabla_\beta {\mathcal N}^\alpha 
= - a({\mathcal N})^\alpha {\mathcal N}_\beta -\omega({\mathcal N})^\alpha{}_\beta +\theta({\mathcal N})^\alpha{}_\beta\,,
\eeq
namely the acceleration $a({\mathcal N})^\alpha=\nabla_{\mathcal N}\, {\mathcal N}^\alpha$  of the congruence, its vanishing vorticity tensor $\omega({\mathcal N})=0$ since the congruence is hypersurface-forming  and the expansion tensor $\theta({\mathcal N})\equiv -K({\mathcal N})$. The scalar expansion ${\rm Tr}\,[\theta({\mathcal N})]=-{\rm Tr}\,[K({\mathcal N})]$ of the congruence is just the sign-reversed mean curvature of the slicing, while the tracefree part of the expansion tensor is the shear tensor.

\end{enumerate}

Finally, there exist slicings  motivated more by analytic considerations than geometrical ones.
One example is represented by the so-called harmonic slicings. In this case the $T=const$ foliation is characterized by a harmonic condition
\beq
\label{harm-cond}
\Box T\equiv \nabla^\mu\partial_\mu T=0\,.
\eeq
A similar condition imposed on all spacetime coordinates specifies the so-called de Donder gauge choice of coordinates.
In the literature, there are also variations of this condition (``$1+\log$" slicing, etc.), which will not be explored here (see, e.g., Ref.~\cite{gourg}).
Eq.~(\ref{harm-cond}) is equivalent to 
\beq
\frac{1}{\sqrt{-g}}\frac{\partial T}{\partial X^\mu}\left(\sqrt{-g}g^{\mu\nu}\frac{\partial T}{\partial X^\nu} \right)=0\,,
\eeq
which becomes in a coordinate system with time coordinate $T=X^0$
\beq
\partial_{\mu}\left(\sqrt{-g}g^{\mu T} \right)=0\,.
\eeq
Recalling Eq.~(\ref{eq:gLL}) the previous equation can be also written as
\beq
-\partial_T \left(\frac{\sqrt{\h }}{L}\right)+\partial_{X^a} \left(\frac{\sqrt{\h}}{L}L^a\right)=0\,,
\eeq
which in turn can be expressed as an evolution equation for the new lapse function $L$.

\section{Minkowski  spacetime}

As a simple example of the problem of finding special slices in a given spacetime, let us consider the flat Minkowski spacetime geometry in an inertial time slicing, with its line element written in spherical coordinates to compare later with a black hole spacetime in Boyer-Lindquist coordinates
\beq
\label{flat_schw}
\rmd s^2 = -\rmd t^2+\rmd r^2 +r^2 \rmd \theta^2 +r^2 \sin^2 \theta\,\rmd \phi^2\ .
\eeq
The $t=const$ hypersurfaces are both intrinsically and extrinsically flat.
Because of their simplicity, the Minkowski slicing examples are useful in better understanding the corresponding general relativistic situations considered below.

Consider a new spherical symmetric slicing by a time function $T=t+f(r)$, retaining the spatial coordinates (and hence the orthogonal time lines).
The induced metric on a typical slice $\Sigma$ is
\beq
(\h_{ab})={\rm diag}\left(1-f_r^2, r^2,r^2\sin^2\theta\right)\,,
\eeq
with the new lapse and shift functions 
\beq
L=\frac1{\sqrt{1-f_r^2}}\,, \qquad
L_a=-f_r\delta^1{}_a\,, \qquad
L^a=-\frac{f_r}{1-f_r^2} \delta^a{}_1\,.
\eeq
One can define the new spatial frame
\beq
E_1=\partial_r +f_r \,\partial_t\,, \qquad
E_2=\partial_\theta\,, \qquad
E_3=\partial_\phi  \,,
\eeq
with dual frame
\beq
\Omega^1=\rmd r\,, \qquad
\Omega^2=\rmd \theta\,, \qquad
\Omega^3=\rmd \phi  \,.
\eeq
In this frame the extrinsic curvature tensor has components
\beq
(K({\mathcal N})^a{}_b)
=-\frac{1}{\sqrt{1-f_r^2}}{\rm diag}\left(\frac{f_{rr}}{1-f_r^2}, \frac{f_r}{r},\frac{f_r}{r}\right)\,,
\eeq
with trace
\beq
\label{eq:TrKr}
{\rm Tr}\,[K({\mathcal N})]
=-\frac{1}{\sqrt{1-f_r^2}}\left(\frac{f_{rr}}{1-f_r^2}+\frac{2f_r}{r}\right)\,.
\eeq
The intrinsic curvature is characterized in three dimensions equivalently either by the Riemann or Ricci tensor, with nonvanishing components respectively
\beq
R(\h)_{r\theta r\theta}=-\frac{rf_r f_{rr}}{1-f_r^2} 
  = \frac{R(\h)_{r\phi r\phi}}{\sin^2 \theta}, \
R(\h)_{\theta\phi\theta\phi}
 =-\frac{r^2 \sin^2 \theta f_r^2}{1-f_r^2}\,,
\eeq
and
\beq
R(\h)^r{}_r
 =-\frac{2f_r f_{rr}}{r(1-f_r^2)}\,,\quad
R(\h)^{\theta}{}_{\theta}
 = -\frac{f_r}{r^2(1-f_r^2)^2}[rf_{rr}+f_r(1-f_r^2)]
=R(\h)^{\phi}{}_{\phi}\,,
\eeq
and the curvature scalar is
\beq\label{eq:Rscalar}
R(\h)
 =\frac{2f_r}{r^2(1-f_r^2)^2}[2rf_{rr}+f_r(1-f_r^2)] \,.
\eeq
Finally, the Cotton-York tensor (\ref{CYtensdef}) is identically zero for this slicing, independent of the choice of $f(r)$, ensuring the conformal flatness of any spherical slicing.

The new time coordinate hypersurfaces $T=const$ have unit normal
\beq
{\mathcal N}^\flat=-L (\rmd t -\nu\,\rmd r)\,, \qquad 
{\mathcal N}=\gamma(\partial_t+\nu\,\partial_r)\,,
\eeq
with radial relative velocity $\nu=f_r$ (which must satisfy $|\nu|<1$) and associated Lorentz factor $\gamma=1/\sqrt{1-\nu^2}$. The associated observers moving orthogonal to the slicing follow radial trajectories which are ingoing for $\nu<0$ and outgoing for $\nu>0$ in comparison with the usual static observers ($\nu=0$) following the original time lines. The simple transformation $f\to-f$ flips the radial motion of the new observers.

Let us consider some explicit examples.
In order to deal with dimensionless quantities, we introduce a positive scaling constant with the dimensions of a length, say $A$.
For simplicity the solutions of the various conditions on the slicing function $f(r)$ below will be given modulo an overall sign subject to the initial condition $f(0)=0$ for the sake of easy comparison.

\begin{enumerate}

\item {\it CMC}:

Let ${\rm Tr}\,[K(\mathcal{N})]=k/A= const$, with $k$ a dimensionless constant.
Eq.~(\ref{eq:TrKr}) then gives
\beq
\label{Xmink}
\frac{f_r}{\sqrt{1-f_r^2}}=-\frac{kr}{3A}+\frac{CA^2}{r^2}\equiv X\,,
\eeq
where $C$ is a dimensionless integration constant, implying that 
\beq
\h_{rr}=\left(1+X^2\right)^{-1}\,, \qquad
\nu=X\sqrt{\h_{rr}}
\,.
\eeq
The function $f(r)$ is then given by 
\beq
f(r)=\int_0^r X\sqrt{\h_{rr}}\,\rmd r\,.
\eeq

\item {\it Maximal}:

These slices are characterized by the vanishing of the trace of the extrinsic curvature tensor  Eq.~(\ref{eq:TrKr}) which leads to
\beq
f_r=\nu=\frac{1}{\sqrt{1+\displaystyle {r^4}/{A^4}}}\,,
\eeq
modulo a constant chosen to be 1 with no loss of generality.
This equation can be integrated in terms of elliptic functions, i.e.,
\beq
f(r)=-\frac{A}{2}\left\{F \left(\arccos\left(\frac{r^2-A^2}{r^2+A^2}\right),\frac1{\sqrt{2}}\right)-2K\left(\frac1{\sqrt{2}}\right)\right\}\,,
\eeq
where $F(x,k)$ and $K(k)$ are the incomplete and the complete elliptic integrals of first kind, respectively (such that $F(\pi,k)=2K(k)$).
Note that in the limit of $r\to\infty$ we have $f(r)\to AK(1/{\sqrt{2}})\approx 1.854\,A$.

\item {\it Vanishing Ricci scalar}:

Requiring that the spatial Ricci scalar Eq.~(\ref{eq:Rscalar}) vanish leads to a parabola of revolution about an inertial time axis orthogonal to its symmetry axis
\beq
f(r)=2\sqrt{A(A+ r)} -2A\,, \quad
\nu=\frac1{\sqrt{1+\displaystyle{r}/{A}}}\,.
\eeq
The induced metric is
\beq
(\h_{ab})={\rm diag}\left(\frac{r}{A+r},r^2,r^2\sin^2\theta\right)\,,
\eeq
with extrinsic curvature
\beq
(K(\mathcal{N})^a{}_b) =\sqrt{\frac{A}{r^3}}\, {\rm diag}\left(\frac{1}{2},-1,-1\right)\,, \quad
{\rm Tr}\,[K(\mathcal{N})]=-\frac32 \sqrt{\frac{A}{r^3}}\,,
\eeq
and intrinsic curvature specified by
\beq
(R(\h)^a{}_b) =\frac{A}{r^3}{\rm diag}\left(1,-\frac12, -\frac12 \right)\,.
\eeq

\item {\it Hyperboloidal}:

Smarr and York \cite{smarr-york} introduced a special constant mean curvature slicing of Minkowski spacetime by translating a single sheet of a spacelike pseudosphere (hyperboloid) of a given fixed radius (and therefore fixed intrinsic and extrinsic curvature) along the time lines of an inertial coordinate system, in contrast with the geodesically parallel family of all such pseudospheres of varying radii centered on one spacetime point.
In the metric (\ref{flat_schw})  in spherical coordinates in an inertial coordinate system, consider the time slices $T=const$ where $T$ is given implicitly by
\beq
\label{Thyp}
(t+A-T)^2-r^2=A^2
\eeq
so that $f(r)=\sqrt{r^2+A^2}-A$, where $A$ is the radius of the pseudosphere and we have chosen only one of the two possible signs, i.e., the plus sign. One finds then the corresponding radial velocity given by
\beq
\nu=\frac1{\sqrt{1+\displaystyle {A^2}/{r^2}}}\,.
\eeq
The normal congruence to the $T=const$ slicing defines the associated Smarr-York observers.
The intrinsic metric of the slices evaluates to
\beq
(\h_{ab})={\rm diag}\left(\frac{A^2}{r^2+A^2},r^2,r^2\sin^2\theta\right)\,,
\eeq
while the extrinsic curvature tensor
\beq
K(\mathcal{N})_{ab} =-\frac{1}{A}\h_{ab}\,,\qquad
{\rm Tr}\,[K(\mathcal{N})]=-\frac{3}{A}\,,
\eeq
and Ricci curvature tensor
\beq
R(\h)_{ab}=\frac{2}{A^2}\h_{ab}\,, \qquad 
R(\h) =\frac{6}{A^2}\,,
\eeq
are both (spacetime) constant multiples of the spatial metric reflecting the constant intrinsic and extrinsic curvature conditions.

The lapse function and shift vector field are given by 
\beq
L=\sqrt{1+\frac{r^2}{A^2}}\,, \qquad
\vec L=L^a\partial_a=-\frac{r}{A}\sqrt{1+\frac{r^2}{A^2}}\, \partial_r\,,
\eeq
respectively.
The spatial metric then takes the form
\beq
\three \rmd s^2=\frac{\rmd r^2}{L^2}+r^2\, \rmd \Omega^2, \qquad 
\rmd \Omega^2
=\rmd \theta^2+\sin^2\theta\, \rmd \phi^2\,.
\eeq
or transforming the radial coordinate by $r=A\sinh \eta$, 
\beq
\three \rmd s^2=A^2(\rmd \eta^2+\sinh^2\! \eta\, \rmd \Omega^2)\,,
\eeq
more familiar from cosmology.
Note that when $A\to \infty$ the slice tends to be both intrinsically and extrinsically flat. This shift vector field satisfies the minimal-distortion equation of Smarr and York.

\end{enumerate}

\begin{figure}
\typeout{*** EPS figure 1}
\begin{center}
\includegraphics[scale=0.3]{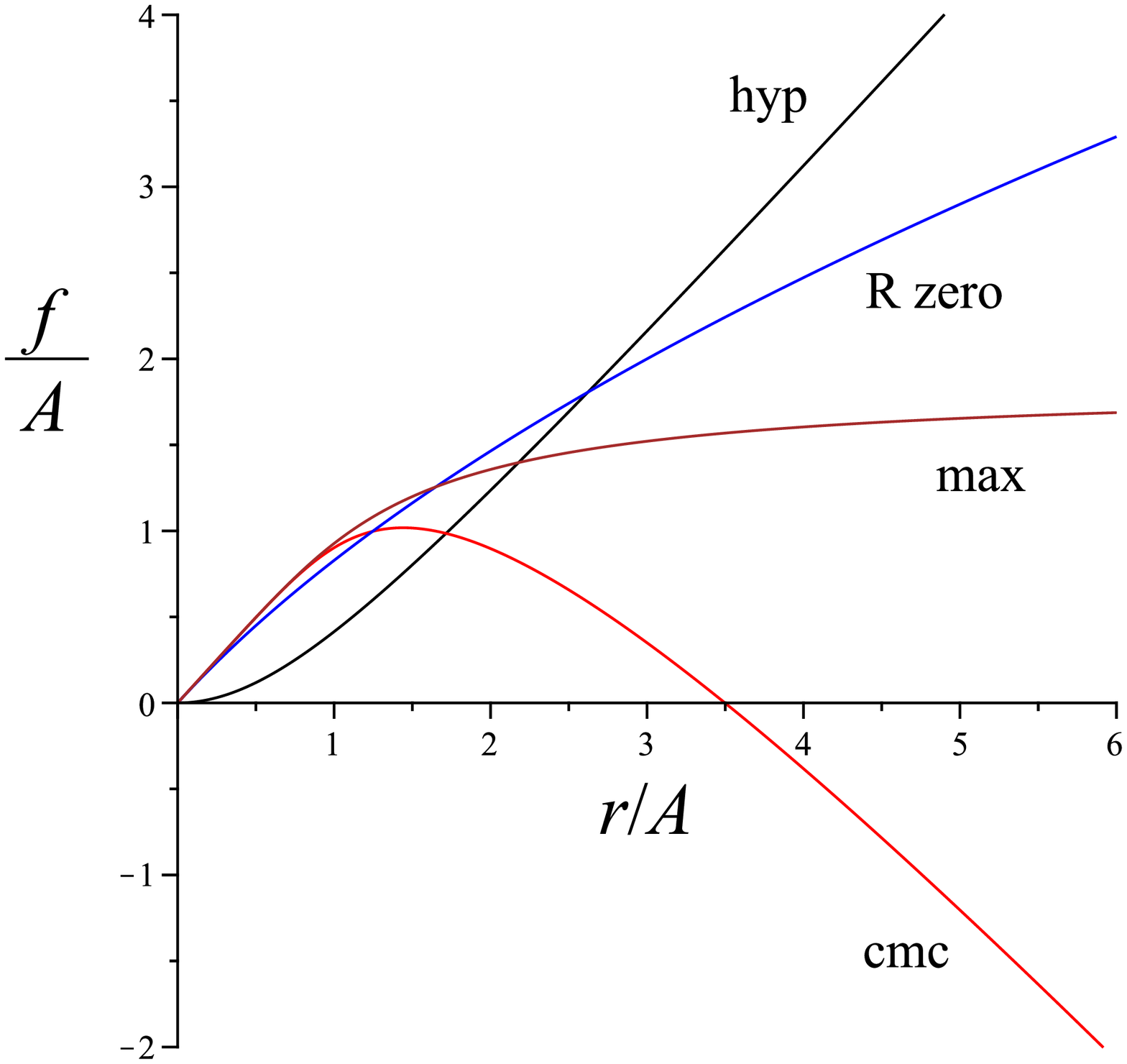}\qquad
\includegraphics[scale=0.3]{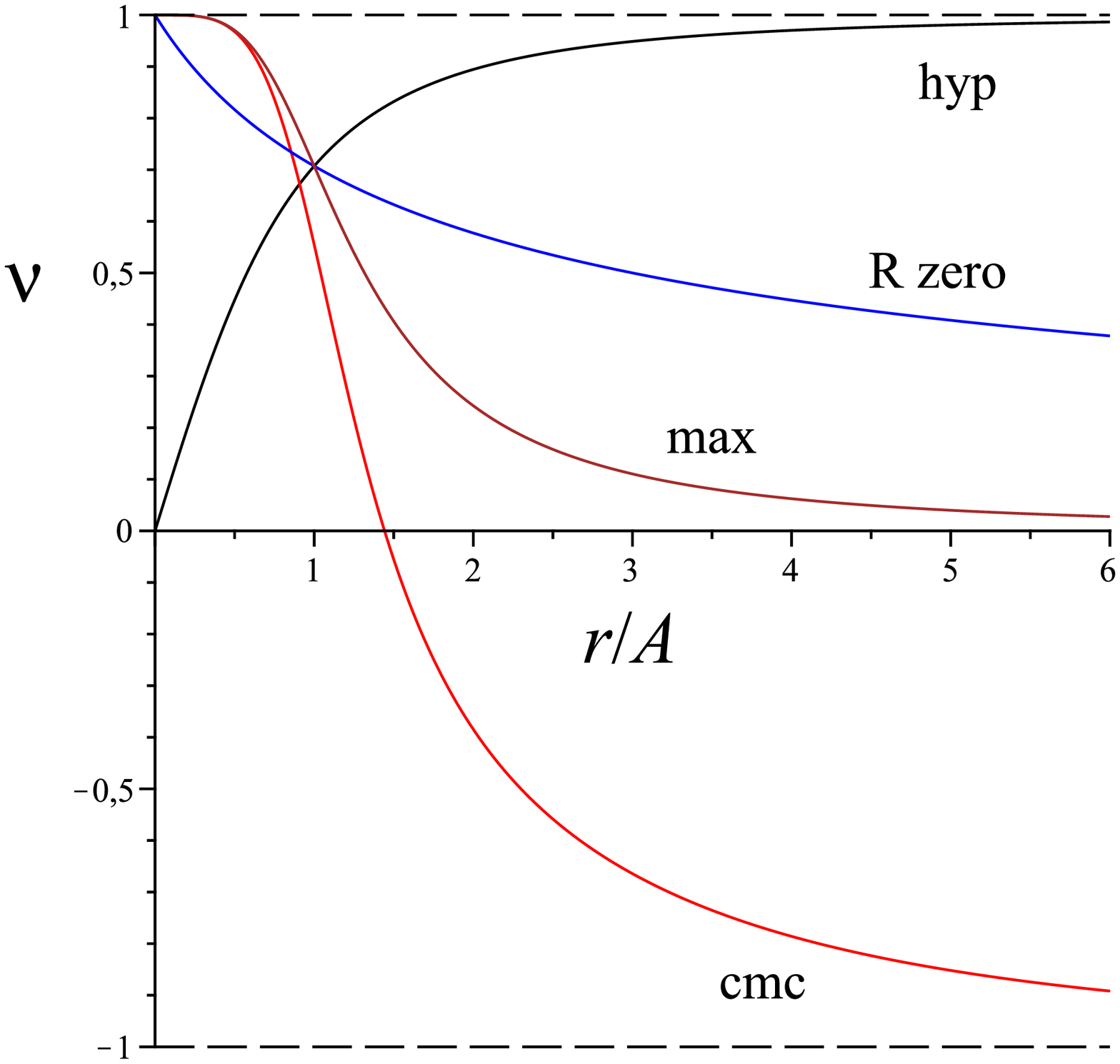}
\end{center}
\caption{Minkowski spacetime. 
The radial behavior of the slicing function $f(r)$ (rescaled by $A$) and the associated spatial velocity $\nu=f_r$ of observers normal to the slicing itself relative to the observers following the coordinate time lines are shown as functions of $r/A$ in all cases discussed in the text: CMC, maximal, vanishing Ricci scalar and hyperboloidal slicings. 
The curve corresponding to the CMC slicing is drawn here for $k=1$; as $k$ approaches $0$, this curve approaches the maximal slicing curve more and more. Note that although all these slicings have observer world lines which are initially radially outgoing at the origin of spatial coordinates, the CMC observers reverse direction to become radually ingoing. Only the hyperboloidal slicing is regular at the origin, while the remaining hypersurfaces have an asymptotically null conical singularity there.
Reversing the sign of $f(r)$ simply interchanges ingoing and outgoing directions for the relative motion of the new observers with respect to the inertial observers of the rest system.
}
\label{fig:1}
\end{figure}

The behavior of the slicing function $f(r)$ as well as of the associated spatial velocity $\nu$ as a function of $r/A$ is shown in Fig.~\ref{fig:1} in all cases discussed above.

% figure 1

\section{Schwarzschild spacetime}

Consider now the Schwarzschild spacetime representing a nonrotating black hole, whose line element written in standard coordinates $(t,r,\theta,\phi)$ is given by
\begin{eqnarray}
\rmd s^2 &=&  -\left(1-\frac{2M}{r}\right)\rmd t^2 +\left(1-\frac{2M}{r}\right)^{-1}\rmd r^2 +r^2(\rmd \theta^2+ \sin^2 \theta \,\rmd \phi^2)\,.
\end{eqnarray}
Introduce an orthonormal frame adapted to the static observers following the time lines, i.e., 
\beq
n=N^{-1}\partial_t\,, \qquad
e_{\hat r}=N\partial_r\,, \qquad
e_{\hat \theta}=\frac{1}{r}\partial_\theta\,, \qquad
e_{\hat \phi}=\frac{1}{r\sin \theta}\partial_\phi\,,
\eeq
with dual frame
\beq
n^\flat=-N\rmd t\ , \quad \omega^{{\hat r}}=N^{-1}\rmd r\ , \quad
\omega^{{\hat \theta}}= r \,\rmd \theta\ , \quad
\omega^{{\hat \phi}}=r\sin \theta\, \rmd \phi\ ,
\eeq
where $N=\sqrt{1-2M/r}$.

The $t=const$ hypersurfaces form a slicing which is extrinsically flat (as the orthogonal hypersurfaces to the static Killing vector congruence $\partial_t$), but not intrinsically flat. 
The induced metric on these hypersufaces
\beq
\three \rmd s^2 =\left(1-\frac{2M}{r}\right)^{-1}\rmd r^2 +r^2(\rmd \theta^2+ \sin^2 \theta \rmd \phi^2)
\eeq
has nonzero Ricci curvature  
\beq
(R(\h)^a{}_{b})=\frac{M}{r^3}\, {\rm diag}\left(-2,1,1\right)\,,
\eeq
with vanishing Ricci scalar $R(\h)=0$.

Let us look for general slicings of the Schwarzschild geometry which are compatible with the Killing symmetries of the spacetime, i.e., spherical slices $T=t+f(r)$.
Their timelike unit normal is
\beq
{\mathcal N}^\flat=-L (\rmd t -f_r\,\rmd r)\,, \qquad 
{\mathcal N}=\gamma (n+\nu\, e_{\hat r})\,,
\eeq
with relative velocity  (see Eq.~(\ref{eq:gammanu})) 
\beq\label{eq:nuf}
\nu\equiv \nu^{\hat r} =N^2f_r\, \leftrightarrow\, f_r = N^{-2} \nu 
\eeq
and associated Lorentz factor $\gamma=1/\sqrt{1-\nu^2}$.
As in the Minkowski spacetime case, reversing the sign of $f(r)$ interchanges the ingoing and outgoing radially moving observers associated with the new slicing.

The induced metric on $\Sigma$ is
\beq
\label{met_ind_sch}
(\h_{ab}) = {\rm diag}\left((\gamma N)^{-2}, g_{\theta\theta}, g_{\phi\phi}\right)\,,
\eeq
and the new lapse function and shift vector field are
\beq
\label{newLschw}
L=N\gamma\,, \quad 
L_a=-\nu\delta^1{}_{ a}\,, \quad 
L^a=-\gamma^2 N^2 \nu\,\delta^{a}{}_1\,.
\eeq
A (nonorthonormal) basis on $\Sigma$ is given by
\beq
E_1=\partial_r +f_r \partial_t\,, \quad
E_2=\partial_\theta\,, \quad
E_3=\partial_\phi\,,
\eeq
with dual frame
\beq
\Omega^1=N\gamma^2 [\nu\, n^\flat + \omega^{\hat r}]\,,\quad 
\Omega^2=\frac{1}{\sqrt{g_{\theta\theta}}}\omega^{\hat \theta}=\rmd \theta\,,\quad 
\Omega^3=\frac{1}{\sqrt{g_{\phi\phi}}}\omega^{\hat \phi}=\rmd \phi\,.
\eeq
The acceleration $a(\mathcal N)$ is given by
\beq
\label{accschw}
a(\mathcal N)=\frac{1}{N\gamma}\partial_r(N\gamma)\,\Omega^1\,,
\eeq
and the extrinsic curvature of the $T=const$ slices then turns out to be 
\beq
K(\mathcal N)=
-\frac{\partial_r X}{N^2\gamma^2}\,\Omega^1\otimes \Omega^1 - rX \left[ \Omega^2\otimes \Omega^2 + \sin^2\theta\,\Omega^3\otimes \Omega^3 \right] \,,
\eeq 
where $X=N\gamma \nu$, and the traces of its powers then have the following values
\beq\label{trKSchw}
{\rm Tr}[K(\mathcal N)^n]=(-1)^n\left[(\partial_r X)^n+\frac{2}{r^n}X^n\right]\,, \quad
(n=1,2,3)\,.
\eeq

The intrinsic curvature is described by the following nonvanishing components of the spatial Riemann tensor
\beq
\label{RMNschw}
R(\h)_{r\theta r \theta} 
=\frac{R(\h)_{r\phi r \phi}}{\sin^2 \theta}
=-\frac{r}{N\gamma}\partial_r(N\gamma)\,, \quad
R(\h)_{\theta \phi \theta \phi}
=r^2\sin^2\theta\,(1-\gamma^2N^2)\,,
\eeq
or equivalently by the spatial Ricci tensor, whose nonzero components are
\beq
R(\h)^r{}_r=-\frac2r N\gamma\partial_r(N\gamma)\,,\quad
R(\h)^\theta{}_\theta=R(\h)^\phi{}_\phi
=\frac12 R(\h)^r{}_r+\frac1{r^2}(1-\gamma^2N^2)\,.
\eeq
Finally, the spatial Ricci scalar is 
\beq
\label{Rschw}
 R(\h)=\frac2{r^2 }\partial_r\left[ r(1-\gamma^2N^2)\right]\,,
\eeq
showing that $r(1-\gamma^2 N^2)=const$ corresponds to vanishing spatial scalar curvature, but $\gamma N=1$ to vanishing spatial curvature.

The Cotton-York tensor (\ref{CYtensdef})  vanishes identically, so that spherical slicings are automatically conformally flat.

Geodesic slicings correspond to a spatially constant lapse function $L = N \gamma=L_0$, as from Eq. (\ref{accschw}).
In this case 
\beq
\label{nugeo}
\gamma=\frac{L_0}{N}\,, \qquad
\nu=\pm\frac1{L_0}\sqrt{L_0^2-N^2}\,,
\eeq
and the only surviving component of the spatial Riemann tensor is 
\beq
R(\h)_{\theta \phi \theta \phi}
=r^2\sin^2\theta\,(1-L_0^2)\,,
\eeq
whereas the spatial Ricci tensor is fully specified by
\beq
R(\h)^\theta{}_\theta=R(\h)^\phi{}_\phi
=\frac1{r^2}(1-L_0^2)\,,
\eeq
with $R(\h)=2(1-L_0^2)/r^2$.
The value of the parameter $L_0$ determines the sign of the intrinsic curvature of the slices, corresponding to either vanishing ($L_0=1$), positive ($L_0<1$) or negative ($L_0>1$) curvature.
Note that interpreting $\mathcal N$ as the 4-velocity field of a family of (geodesic) test particles, the parameter $L_0$ coincides with the (conserved, Killing) energy per unit mass of the particles
\beq
\tilde E=-{\mathcal N}_t=N\gamma=L_0\,.
\eeq
Therefore, $L_0=1$ also represents the case of particles with vanishing radial velocity at spatial infinity; $L_0>1$ can be associated with particles starting at infinity with nonzero (inward) velocity; finally, $L_0<1$ corresponds to particles starting moving at a finite radial position.
In the literature, adapted coordinates to these situations (in the special case of a Schwarzschild spacetime) are referred to as \lq\lq hail'' ($L_0>1$), \lq\lq rain'' ($L_0=1$) and \lq\lq drip'' ($L_0<1$) coordinates (see, e.g., Ref. \cite{Finch} and references therein).

\begin{figure}
\typeout{*** EPS figure 2}
\begin{center}
\includegraphics[scale=0.3]{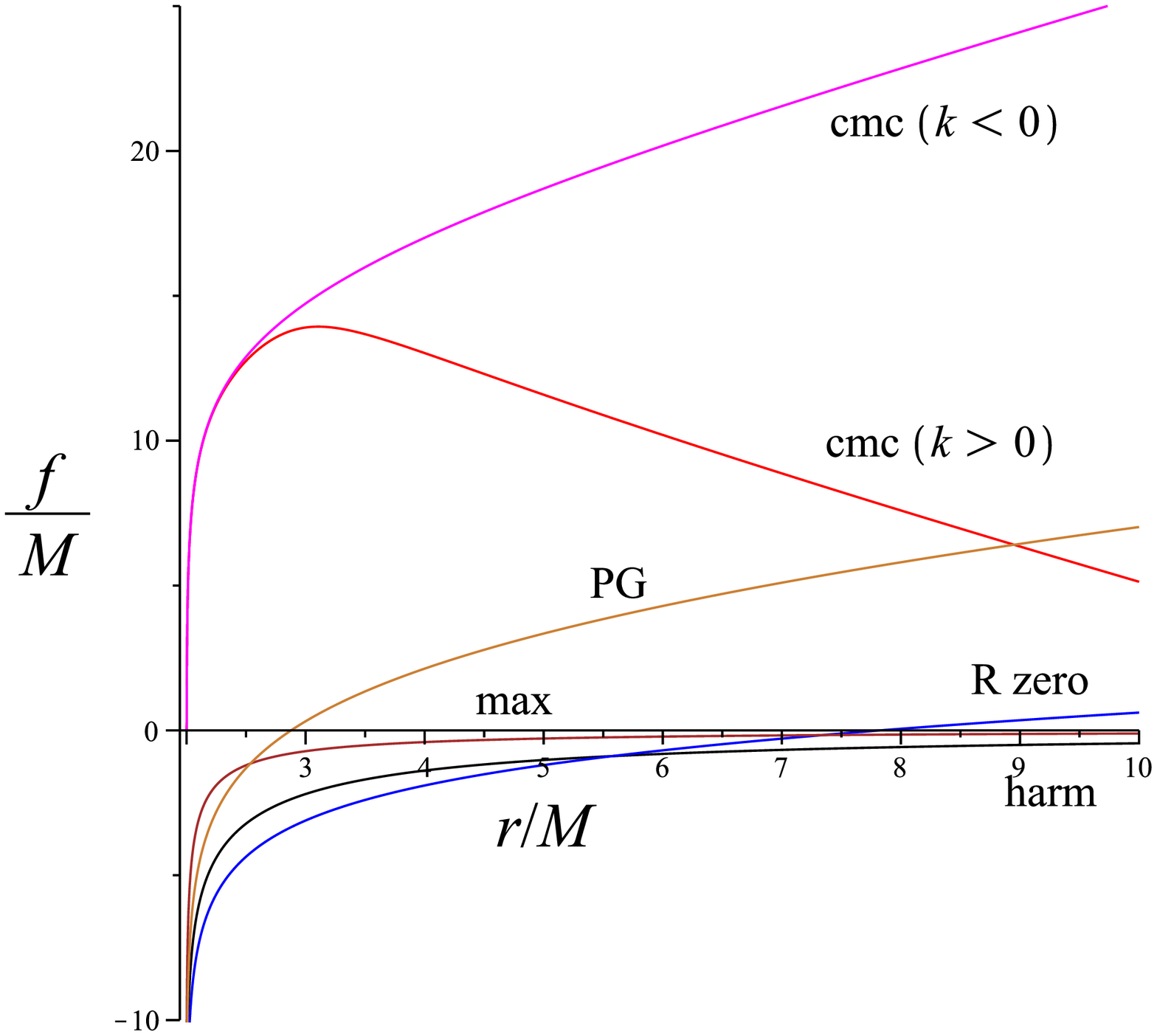}\qquad
\includegraphics[scale=0.3]{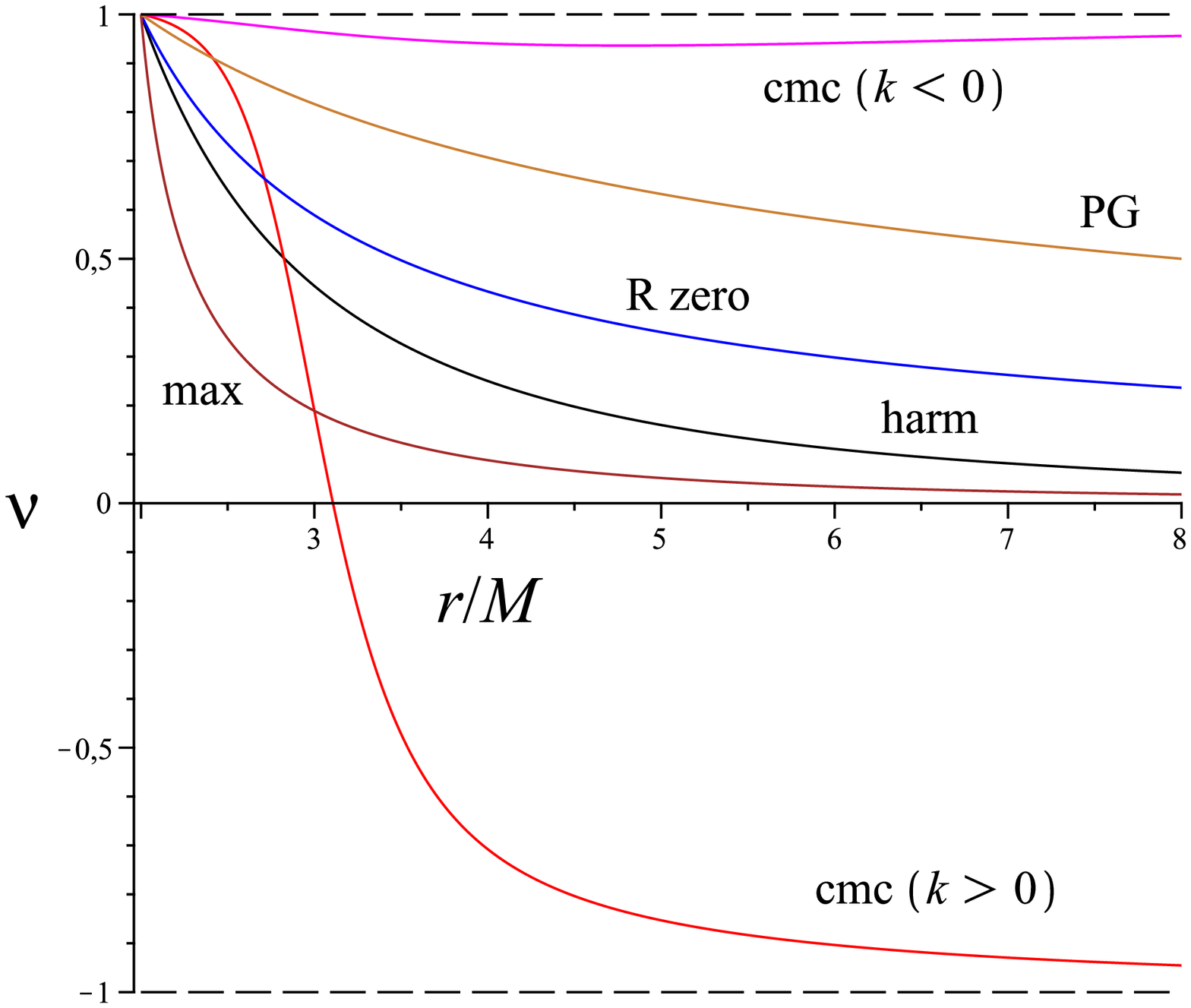}
\end{center}
\caption{Schwarzschild spacetime.
The comparative behavior of the slicing function $f(r)$ (left) and the corresponding radial velocity $\nu$ (right) associated with the special observer families considered in the text (i.e., CMC, maximal, vanishing Ricci scalar, harmonic and PG) is shown for the following choice of parameters: $C=10$, $k=\pm1$ (CMC); $C=10$ (maximal); $C=1$ (vanishing Ricci scalar); $C=4$ (harm). The outgoing PG slicing ($\nu>0$) is shown to compare with the remaining mostly outgoing radial coordinate associated observers, apart from the positive curvature CMC case which has  both radially ingoing and outgoing regimes of their associated observers compared to the usual static observers to which all of these are compared ($\nu=0$). Note that $f$ diverges at radial infinity for the zero curvature, PG, and constant mean curvature slicings, and at $r=2M$ for all but the constant mean curvature slicings, while the maximal and harmonic slicings are chosen to vanish at radial infinity. 
}
\label{fig:2}
\end{figure}

Let us consider some explicit examples.
We will let $C$ denote in each case the dimensionless integration constant appearing in the solution of the ordinary differential equation for the radial velocity $\nu$, all quantities being rescaled by the characteristic length scale of the background curvature (i.e., the mass $M$).
For simplicity the solutions for the slicing function $f(r)$ will be given up to an overall sign and the constant of integration will be chosen so that $f(2M)=0$ if this function is finite at $r=2M$, or $\lim_{r\to\infty}f(r)=0$ if it approaches a finite limit at $r\to\infty$, in order to compare the time slices from a common reference point at the horizon or at spatial infinity relative to the usual slicing.
Typical behaviors of the slicing functions as well as of the corresponding linear velocities are shown in Fig.~\ref{fig:2}.

\begin{enumerate}

\item {\it CMC}:

Let $M\,{\rm Tr}\,[K(\mathcal{N})]=k=const$. Then using
Eq.~(\ref{trKSchw}) with $n=0$ one finds
\beq
X=-\frac{kr}{3M}+\frac{CM^2}{r^2}\,,
\eeq
formally equivalent to Eq.~(\ref{Xmink}) in the case of a flat spacetime, where $C$ is a dimensionless integration constant, implying that 
\beq
\nu=\frac{X}{\sqrt{N^2+X^2}}\,,
\eeq
and
\beq
\h_{rr}=\frac{\nu^2}{X^2}=(N^2+X^2)^{-1}\,.
\eeq
The function $f(r)$ is then given by 
\beq
f(r)=\int_{2M}^r\frac{X}{N^2\sqrt{N^2+X^2}}\,\rmd r\,.
\eeq
Let $k>0$. Then the radial velocity approaches $\nu\to-1$ for large $r$, independent of the value of $C$.
In contrast the behavior at the horizon depends on the sign of the quantity $-8k+3C$. In fact, in the limit $r\to2M$ one has $\nu\to{\rm sgn}(-8k+3C)$, i.e., $\nu\to1$ if $C>\frac83k$, while $\nu\to-1$ if $C<\frac83k$.

\item {\it Maximal}:

For $k=0$ the above relations simplify to
\beq
X=\frac{CM^2}{r^2}\,,
\eeq
so that $\nu=CM^2{\sqrt{\h_{rr}}}/r^2$ and
\beq
\h_{rr}=\left(1-\frac{2M}{r}+\frac{C^2M^4}{r^4}\right)^{-1}\,,
\eeq
and
\beq
f(r)=-CM^2\int_r^{\infty}\frac{\sqrt{\h_{rr}}}{r^2N^2}\,\rmd r\,,
\eeq
which can be expressed in terms of elliptic functions.
Therefore the radial velocity behaves as $\nu\to0$ for large $r$ and $\nu\to1$ for $r\to2M$, independent of the value of $C$.

\item {\it Vanishing Ricci curvature}:

Setting $R(\h)=0$ in Eq.~(\ref{Rschw}) gives $r(1-\gamma^2 N^2)=const$, leading to
\beq
\nu=\frac{1}{\sqrt{1+C N^2 r/M}}\,,
\eeq
and
\beq
f(r)=\frac{M}{C\nu}(1-\nu)+2M\ln\left(\frac{1-\nu}{1+\nu}\right)\,.
\eeq
Therefore the radial velocity behaves as $\nu\to0$ for large $r$ and $\nu\to1$ for $r\to2M$, independent of the value of $C$.

\item {\it Harmonic}:

The harmonic condition (\ref{harm-cond}) gives
\beq
0=\partial_r\nu+\frac{2\nu}{r}\,,
\eeq
so that $\nu=CM^2/r^2$ and
\beq
f(r)=CM\ln N\,.
\eeq
$C=4$ is a common choice in the literature.
Therefore the radial velocity behaves as $\nu\to0$ for large $r$ and $\nu\to1$ for $r\to2M$ (for that choice of the integration constant).

\end{enumerate}

Let us investigate now spherical slicings characterized by intrinsic flatness, i.e., by the vanishing of the spatial Riemann tensor. Examining Eq.~(\ref{RMNschw}) it is sufficient to impose the condition 
\beq
L=1 \leftrightarrow\gamma=\frac{1}{N}\,,\qquad
\nu=\pm\sqrt{1-N^2}\,,
\eeq
which obviously yields a flat induced metric apparent from Eq.~(\ref{met_ind_sch}) since $\h_{rr}=1$, while similarly the acceleration vanishes as well from Eq.~(\ref{accschw}), leading to a geodesically parallel slicing.  
Because of the above choice of the lapse function, the slicing is associated with the \lq\lq rain'' coordinates introduced above, whose adapted observers are known as Painlev\'e-Gullstrand (PG) observers \cite{Painleve21,Gullstrand22}. They are geodesic and irrotational, and admit orthogonal time hypersurfaces which are intrinsically flat. 
Their radial velocity relative to the static observers is $\nu=-\sqrt{1-N^2}=-\sqrt{2M/r}$, having selected the minus sign corresponding to a free fall from rest at infinity (ingoing PG, for example).
Integrating Eq.~(\ref{eq:nuf}) then gives
\beq
f(r)=
-2\sqrt{2Mr}+2M\ln\frac{\sqrt{r}+\sqrt{2M}}{\sqrt{r}-\sqrt{2M}}\,.
\eeq
However, PG slicings are not extrinsically flat since
\beq
M\,K(\mathcal{N})= \sqrt{\frac{2M^3}{r^3}} 
\left[ \frac12  \Omega^1\otimes \Omega^1 -  r^2\,\Omega^2\otimes \Omega^2 - r^2\sin^2\theta\,\Omega^3\otimes \Omega^3\right]\,.
\eeq
PG observers are in a sense \lq\lq complementary" to the static observers, which instead have extrinsically flat orthogonal time hypersurfaces.

\section{Splitting of the curvature tensor by a generic family of observers}

For rotating black holes, the situation is more complicated, so we examine first the splitting of the curvature tensor using the test observer congruence associated with a given stationary spacelike slicing.
Let $U$ be the unit timelike $4$-velocity vector tangent to the test observer world lines orthogonal to the slicing. 
The orthogonal decomposition of the covariant derivative of this irrotational congruence $U$ is
\beq
\nabla_\alpha U^\beta =-U_\alpha a(U)^\beta  -\theta(U)^\beta{}_\alpha\,,
\eeq
where $a(U)^\beta=U^\alpha\nabla_\alpha U^\beta$ is the acceleration and 
\beq
\theta(U)_{\alpha\beta}=P(U)^\mu_\alpha P(U)^\nu_\beta\nabla_{(\mu}U_{\nu)}
\eeq
is the expansion tensor with trace $\Theta(U)=\theta(U)^\alpha{}_\alpha$, the expansion scalar usually referred to as just the expansion of the congruence, while the tracefree part of the expansion tensor $\sigma(U)_{\alpha\beta}$ describes the shear of the congruence.

Details of the decomposition of the curvature tensor and corresponding Einstein equations may be found in many places, including Ref.~\cite{mfg}. 
Using an observer-adapted frame, the latter lead to the following set of equations in vacuum, setting the Ricci tensor to zero

\begin{eqnarray}
0&=&R^\top{}_\top
=[\nabla(U)_c +a(U)_c]a(U)^c-\nabla(U)_{\rm (lie)}\Theta(U) -{\rm Tr}\,[\theta(U)^2]
\,,\nonumber\\
0&=&R^\top{}_a
=-\nabla(U)_c [\theta(U)^c{}_a-\Theta(U)\delta^c{}_a]
\,,\nonumber\\
0&=&R^a{}_b
=-[\nabla(U)_{\rm (lie)}+\Theta(U)]\theta(U)^a{}_b-[\nabla(U)_{(b}+a(U)_{(b}]a(U)^{a)}+R(U){}^a{}_b
\,,
\end{eqnarray}
where the index $\top$ corresponds to the component tangential to the congruence $U$ (and proportional to the orthonormal temporal component). 
Here $\nabla(U)=P(U)\nabla$ and $\nabla(U)_{\rm (lie)}=P(U)\pounds_U$ denote the spatial covariant derivative and the spatial Lie derivative, respectively.

Introducing the spatial symmetric ``strain" tensor \cite{fdf1,fdf2,fdf3,strains,jets},
\beq
S(U)_{ab}=[\nabla(U)_{(b}+a(U)_{(b}]a(U)_{a)}\,,
\eeq
these equations can be written
\begin{eqnarray}
\label{general_acc}
&&\nabla(U)_{\rm (lie)}\Theta(U)= {\rm Tr}\,[S(U)] -{\rm Tr}[\theta(U)^2]\,,\nonumber\\
&& \nabla(U)_a [\theta(U)^a{}_c-\Theta(U)\delta^a{}_c]=0\,,\nonumber\\
&&R(U){}^a{}_b=[\nabla(U)_{\rm (lie)}+\Theta(U)]\theta(U)^a{}_b+S(U)^a{}_b\,,
\end{eqnarray}
If $U$ is a geodesic congruence (i.e., $a(U)=0$), the strain tensor then vanishes identically  and Eq.~(\ref{general_acc}) reduces to
\begin{eqnarray}
\label{geosli}
&&\nabla(U)_{\rm (lie)}\Theta(U)=  -{\rm Tr}[\theta(U)^2]\,,\nonumber\\
&& \nabla(U)_a [\theta(U)^a{}_c-\Theta(U)\delta^a{}_c]=0\,,\nonumber\\
&&R(U){}^a{}_b=[\nabla(U)_{\rm (lie)}+\Theta(U)]\theta(U)^a{}_b\,.
\end{eqnarray}
Therefore in vacuum for  irrotational geodesic slicings one has the general result that a nonzero extrinsic curvature implies a nonzero spatial Ricci curvature. This is exactly what happens in the Kerr case when considering Painlev\'e-Gullstrand observers.
What is really special in the Schwarzschild case, instead, is that the extrinsic curvature associated with  Painlev\'e-Gullstrand slicings is nonzero but  
\beq
[\nabla(U)_{\rm (lie)}+\Theta(U)]\theta(U)^a{}_b=0\,,
\eeq
implying that $R(U){}^a{}_b=0$.
We will discuss these as well as other related properties in the next section for a Kerr spacetime.

\section{Kerr spacetime}

Consider the Kerr spacetime representing a rotating black hole with its line element written in standard Boyer-Lindquist coordinates $(t,r,\theta,\phi)$ as
\beq 
\rmd s^2 = -\left(1-\frac{2Mr}{\Sigma}\right)\rmd t^2 -\frac{4aMr}{\Sigma}\sin^2\theta\,\rmd t\,\rmd\phi+ \frac{\Sigma}{\Delta}\,\rmd r^2 +\Sigma\,\rmd \theta^2+\frac{\Lambda}{\Sigma}\sin^2 \theta \,\rmd \phi^2\ ,
\eeq
where $\Delta=r^2-2Mr+a^2$, $\Sigma=r^2+a^2\cos^2\theta$ and $\Lambda = (r^2+a^2)^2-\Delta a^2\sin^2 \theta$.
Here $M$ and $a$ with $|a|\le M$ are the total mass and the specific angular momentum of the source respectively. The event horizons are located at $r_\pm=M\pm\sqrt{M^2-a^2}$.

The Boyer-Lindquist $t=const$ slicing of the Kerr spacetime is a well known maximal slicing associated with the zero angular momentum observer (ZAMO) family of fiducial observers, with 4-velocity field orthogonal to the slicing leaves given by 
\beq
\label{n}
n^\flat =-N\rmd t, \qquad n=N^{-1}(\partial_t-N^{\phi}\partial_\phi)\,,
\eeq
where $N=(-g^{tt})^{-1/2}=\left[\Delta\Sigma/\Lambda\right]^{1/2}$ and $N^{\phi}=g_{t\phi}/g_{\phi\phi}=-2aMr/\Lambda$ are the lapse function and the only nonvanishing component of the shift vector field, respectively, satisfying the relations (useful in manipulating expressions)
\beq
g_{\phi\phi}N^2=\Delta \sin^2\theta, \quad  
\frac{N^\phi}{N^2}=-\frac{2aMr}{\Delta \Sigma}\ .
\eeq
An orthonormal frame adapted to the ZAMOs is given by
\beq
\label{zamoframe}
e_{\hat t}=n\ , \quad
e_{\hat r}=\frac1{\sqrt{g_{rr}}}\partial_r=\partial_{\hat r}\ , \quad
e_{\hat \theta}=\frac1{\sqrt{g_{\theta \theta }}}\partial_\theta=\partial_{\hat \theta}\ , \quad
e_{\hat \phi}=\frac1{\sqrt{g_{\phi \phi }}}\partial_\phi=\partial_{\hat \phi}\ ,
\eeq
with dual frame
\beq
\omega^{{\hat t}}=N\rmd t\ , \quad \omega^{{\hat r}}=\sqrt{g_{rr}}\,\rmd r\ , \quad
\omega^{{\hat \theta}}= \sqrt{g_{\theta \theta }} \,\rmd \theta\ , \quad
\omega^{{\hat \phi}}=\sqrt{g_{\phi \phi }}\,(\rmd \phi+N^{\phi}\rmd t)\ .
\eeq

The ZAMOs are accelerated and locally nonrotating, and have a nonzero but tracefree expansion tensor completely described in terms of the (shear) expansion vector $\theta(n)_{\hat \phi}{}^\alpha = \theta(n)^\alpha{}_\beta\,{e_{\hat\phi}}^\beta$ as
\beq
\theta(n)=e_{\hat\phi}\otimes\vec\theta(n)_{\hat\phi}+\vec\theta(n)_{\hat\phi}\otimes e_{\hat\phi}\ .
\eeq
The extrinsic curvature $K(n)=-\theta(n)$ is nonzero, whereas its trace vanishes making the slicing a maximal one. 
${\rm Tr}\,[K(n)^3]$ also vanishes, whereas ${\rm Tr}\,[K(n)^2]$ is nonzero.
In terms of the dimensionless inverse radius $u=M/r$ and the dimensionless rotational parameter $\hat a=a/M$, the latter turns out to be
\beq
M^2 \,{\rm Tr}\,[K(n)^2] =2 {\hat a}^2 u^6 \sin^2(\theta)\frac{ \alpha \tilde\Sigma ^2-4\beta \tilde\Sigma +8 u (1+{\hat a}^2u^2) }{[\tilde\Sigma\tilde\Delta +2 u (1+{\hat a}^2u^2)]^2\tilde\Sigma^3}\,, 
\eeq
with
\beq
\alpha=  -3+8 u-6 {\hat a}^2 u^2 +{\hat a}^4u^4\,,\qquad
\beta = -3+4u-3 {\hat a}^2u^2+2 {\hat a}^2 u^3\,,
\eeq
and
\beq
\tilde \Sigma=1+\hat a^2 u^2 \cos^2\theta\,, \qquad
\tilde\Delta=1-2u+\hat a^2 u^2\,.
\eeq
In the weak field limit $u\ll1$ the previous expression becomes 
\beq
\label{tr_K_quad_zamo}
M^2 \,{\rm Tr}\,[K(n)^2]
=18 \sin^2 \theta \,\hat a^2 u^6 \left[ 1-\frac13  (4+13\cos^2\theta) \hat a^2 u^2\right]+O(u^{10})\,.
\eeq

Similarly, the intrinsic curvature of the Boyer-Lindquist time coordinate slices is nonzero. In fact, the spatial Ricci tensor has the approximated expression
\begin{eqnarray}
\label{eq:6.16}
M^2\,\, (R(\h)^a{}_b) &=& u^3 
\begin{pmatrix}
-2 & 0 & 0\cr
0 & 1 & 0 \cr
0 & 0 & 1 
\end{pmatrix}
+9 \hat a^2\cos\theta\sin\theta\, u^4 
\begin{pmatrix}
0 & 1 & 0\cr
0 & 0 & 0 \cr
0 & 0 & 0 
\end{pmatrix}
\nonumber\\
&+ & 3 \hat a^2 u^5 
\begin{pmatrix}
5\cos^2(\theta)-1  & -6 \cos\theta\sin\theta  & 0\cr
0 &  1-3\cos^2\theta   & 0 \cr
0 & 0 & -2\cos^2\theta
\end{pmatrix}
+O(u^6)\,.
\end{eqnarray}
The three matrices appearing here have zero trace and only at the next order $u^6$ do nonzero trace terms appear. 
In fact, $M^2\, R(\h)=M^2 \,{\rm Tr}\,[K(n)^2]$, which results from combining the first and the third of the general relations (\ref{general_acc}) when $\Theta(U)=0$,  and since $M^2 \,{\rm Tr}\,[K(n)^2]$ starts as $O(u^6)$ in its asymptotic expansion (see Eq.~(\ref{tr_K_quad_zamo})). 

The ZAMO kinematical quantities only have nonzero components in the $r$-$\theta$ plane of the tangent space, i.e.
\begin{eqnarray}
\label{accexp}
a(n) & = & a(n)^{\hat r}\, e_{\hat r}+a(n)^{\hat\theta}\, e_{\hat \theta}
=\nabla(n)\ln N
\,,\nonumber\\
\vec\theta(n)_{\hat\phi} & = &\theta(n)_{\hat\phi}{}^{\hat r}\, e_{\hat r}+\theta(n)_{\hat\phi}{}^{\hat\theta}\, e_{\hat \theta}
=-\frac{\sqrt{g_{\phi\phi}}}{2N}\nabla(n)N^\phi\,.
\end{eqnarray}
For later use it is convenient to introduce the quantity
\beq
\nu_{n}=\sqrt{g_{\phi\phi}}\,\frac{N^\phi}{N}\,, \qquad
\gamma_n=\frac{N}{\sqrt{-g_{tt}}}\,,
\eeq
as well as the curvature vectors $k(x^i,n)$ associated with the diagonal metric coefficients \cite{mfg,idcf1,idcf2,bjdf} 
\beq
k(x^i,n)=k(x^i,n)^{\hat r} e_{\hat r} + k(x^i,n)^{\hat \theta} e_{\hat \theta}
=-\nabla(n)\ln\sqrt{g_{ii}}\,.
\eeq
The relevant ZAMO kinematical quantities are listed in Appendix A.
A recent review of ZAMO slicings can be found in Ref.~\cite{Frolov:2014dta}.

Let us focus on axisymmetric slices $f=f(r,\theta)$, because the Kerr metric is axially symmetric.
The timelike unit normal to these slices is given by
\beq
\label{nuegamma}
{\mathcal N}^\flat
=-L(\rmd t +f_r\rmd r+f_\theta\rmd\theta)\,, \qquad 
{\mathcal N}=\gamma (n+\nu^{\hat r} e_{\hat r}+\nu^{\hat \theta} e_{\hat \theta})\,,
\eeq
with relative velocity components and associated Lorentz factor
\beq
\nu_{\hat a}=\frac{N}{\sqrt{g_{aa}}}f_a\ , \qquad 
\gamma=(1-\delta^{\hat a\hat b}\nu_{\hat a}\nu_{\hat b})^{-1/2}\,.
\eeq
The induced metric on $\Sigma$ is given by
\beq
(\h_{ab})=\begin{pmatrix}
g_{rr}&0&0 \cr
0& g_{\theta\theta}& 0 \cr
0&0& g_{\phi\phi}\cr
\end{pmatrix}
+g_{t\phi}
\begin{pmatrix}
0&0&f_r \cr
0& 0& f_\theta \cr
f_r&f_\theta& 0\cr
\end{pmatrix}
+g_{tt}
\begin{pmatrix}
f_r^2&f_rf_\theta&0 \cr
f_rf_\theta& f_\theta^2& 0 \cr
0& 0&0\cr
\end{pmatrix}\,,
\eeq
or explicitly 
\begin{eqnarray}
\h_{rr}&=& g_{rr}\left( 1-\frac{\nu_{\hat r}^2}{\gamma_n^2}\right), \quad
\h_{\theta\theta}= g_{\theta\theta}\left(  1-\frac{\nu_{\hat \theta}^2}{\gamma_n^2}\right), \quad
\h_{\phi\phi}= g_{\phi\phi} , \nonumber \\
\h_{r\theta}&=& -\frac{\sqrt{g_{rr}g_{\theta\theta}}}{\gamma_n^2}\,\nu_{\hat r}\nu_{\hat \theta}, \quad
\h_{r\phi}=\sqrt{g_{rr}g_{\phi\phi}}\,\nu_n \nu_{\hat r}, \quad
\h_{\theta\phi}= \sqrt{g_{\theta\theta}g_{\phi\phi}}\,\nu_n \nu_{\hat \theta}\,.
\end{eqnarray}
The new lapse is $L=\gamma N$, since $N^cf_c=0$, taking into account that $N_a=N_\phi \delta^\phi{}_a$, whereas the new shift is specified either by the covariant components
\beq
L_1=g_{tt}f_r=-\frac{N\sqrt{g_{rr}}}{\gamma_n^2}\nu_{\hat r}\,, \quad
L_2=g_{tt}f_\theta=-\frac{N\sqrt{g_{\theta\theta}}}{\gamma_n^2}\nu_{\hat \theta}\,, \quad 
L_3=g_{t\phi}\equiv N_\phi\,, 
\eeq
or the contravariant components 
\begin{eqnarray}
\label{Lup}
L^1&=&-\gamma^2 N \frac{\nu_{\hat r}}{\sqrt{g_{rr}}}\,, \quad
L^2=-\gamma^2 N \frac{\nu_{\hat \theta}}{\sqrt{g_{\theta\theta}}}\,, \quad
L^3= \gamma^2 N^\phi\,.
\end{eqnarray}
Finally one can evaluate the (nonorthogonal) basis $E_c$ and its dual frame $\Omega^c$ using Eqs.~(\ref{Lup}), i.e.,
\begin{eqnarray}
E_1=\partial_r +f_r \partial_t, \quad
E_2=\partial_\theta +f_\theta \partial_t, \quad
E_3=\partial_\phi \,,
\end{eqnarray}
and
\begin{eqnarray}
\Omega^1=\rmd r +\frac{\gamma \nu_{\hat r}}{\sqrt{g_{rr}}}\,{\mathcal N}^\flat, \quad
\Omega^2=\rmd \theta +\frac{\gamma \nu_{\hat \theta}}{\sqrt{g_{\theta\theta}}}\,{\mathcal N}^\flat, \quad
\Omega^3=\rmd \phi -\frac{\gamma \nu_n}{\sqrt{g_{\phi\phi}}}\,{\mathcal N}^\flat\,.
\end{eqnarray}

Having written the new spatial metric and the normal congruence ${\mathcal N}$, obtaining both the kinematical quantities of ${\mathcal N}$ as well as the extrinsic and intrinsic curvature of the slice $\Sigma$ is now straightforward.
The 4-acceleration $a({\mathcal N})=\nabla_{\mathcal N}{\mathcal N}$ turns out to be
\beq
a({\mathcal N})=a({\mathcal N})_1\Omega^1+a({\mathcal N})_2\Omega^2\,,
\eeq
with
\begin{eqnarray}
a({\mathcal N})_1&=&\sqrt{g_{rr}}(\partial_{\hat r}\ln L)
=\sqrt{g_{rr}}\left[
a(n)_{\hat r}+\gamma^2(\nu_{\hat r}\partial_{\hat r}\nu_{\hat r}+\nu_{\hat \theta}\partial_{\hat r}\nu_{\hat \theta})
\right]\,, \nonumber\\
a({\mathcal N})_2&=&\sqrt{g_{\theta\theta}}(\partial_{\hat \theta}\ln L)
=\sqrt{g_{\theta\theta}}\left[
a(n)_{\hat \theta}+\gamma^2(\nu_{\hat r}\partial_{\hat \theta}\nu_{\hat r}+\nu_{\hat \theta}\partial_{\hat \theta}\nu_{\hat \theta})
\right]\,.
\end{eqnarray}

With  respect to the spatial frame $\{E_a\}$, the components of the extrinsic curvature tensor 
$K(\mathcal{N})=K(\mathcal{N})_{ab}\Omega^a\otimes \Omega^b$ are given by
\begin{eqnarray}
\label{Kaxisym}
K(\mathcal{N})_{11}&=& g_{rr}\gamma \left[ 
\nu_{\hat r}(W_{\hat r}-2\nu_nZ_{\hat r})+\nu_{\hat r|\hat r}
\right]\,,\nonumber \\
K(\mathcal{N})_{12}&=& -\sqrt{g_{rr}g_{\theta\theta}}\gamma \left[ 
\nu_{\hat r}(W_{\hat \theta}- \nu_n Z_{\hat \theta})-\nu_{\hat \theta}\nu_n Z_{\hat r}
+ \nu_{\hat \theta| \hat r}
\right]\,,\nonumber \\
K(\mathcal{N})_{13}&=& \sqrt{g_{rr}g_{\phi\phi}}\gamma Z_{\hat r}\,,\nonumber\\
K(\mathcal{N})_{22}&=& g_{\theta\theta}\gamma \left[ 
\nu_{\hat \theta}(W_{\hat \theta}-2\nu_n Z_{\hat \theta})+\nu_{\hat \theta |\hat \theta}
\right]\,,\nonumber \\
K(\mathcal{N})_{23}&=& \sqrt{g_{\theta\theta}g_{\phi\phi}}\gamma Z_{\hat \theta}\,,\nonumber\\
K(\mathcal{N})_{33}&=&
-g_{\phi\phi}\gamma\nu_{\hat \phi|\hat \phi}\,,
\end{eqnarray}
where 
\begin{eqnarray}
Z_{\hat a}&=&\nu_{\hat a}\nu_n(k(\phi,n)\cdot\nu)-[\theta(n)_{\hat \phi}{}_{\hat a}- (\theta(n)_{\hat \phi}\cdot \nu )\nu_{\hat a}]\,,\nonumber\\ 
W_{\hat a}&=&\nu_{\hat a}\nu_n^2(k(\phi,n)\cdot\nu)+[a(n)_{\hat a}- (a(n)\cdot \nu)\nu_{\hat a}]\,,
\end{eqnarray}
and $\nu_{\hat a|\hat b}=P(n)^\alpha{}_{\hat a}P(n)^\beta{}_{\hat b}\nabla_\beta \nu_{\alpha}$, i.e.,
\begin{eqnarray}
\nu_{\hat r|\hat r}&=& \partial_{\hat r}\nu_{\hat r}-k(r,n)_{\hat \theta}\nu_{\hat \theta}\,,\qquad
\nu_{\hat r|\hat \theta}= \partial_{\hat \theta}\nu_{\hat r}+k(\theta,n)_{\hat r}\nu_{\hat \theta}\,,\nonumber \\
\nu_{\hat \theta|\hat r}&=& \partial_{\hat r}\nu_{\hat \theta}+k(r,n)_{\hat \theta}\nu_{\hat r}\,,\qquad
\nu_{\hat \theta|\hat \theta}= \partial_{\hat \theta}\nu_{\hat \theta}-k(\theta,n)_{\hat r}\nu_{\hat r}\,,\nonumber \\
\nu_{\hat \phi|\hat \phi}&=& -k(\phi,n)\cdot\nu\,.
\end{eqnarray}
The trace then turns out to be 
\begin{eqnarray}
{\rm Tr}\,[K(\mathcal{N})]&=&-\gamma\left\{
\nu_{\hat \theta|\hat \theta}+\frac{\nu_{\hat \phi|\hat \phi}}{\gamma_n^2}+\gamma^2\left[
\nu_{\hat r|\hat r}+\nu_n^2\nu_{\hat \phi|\hat \phi}+\nu_{\hat r}W_{\hat r}+\nu_{\hat \theta}W_{\hat \theta}\right.\right.\nonumber\\
&&\left.\left.
+(\nu_{\hat r|\hat \theta}+\nu_{\hat \theta|\hat r})\nu_{\hat r}\nu_{\hat \theta}
+(\nu_{\hat \theta|\hat \theta}-\nu_{\hat r|\hat r})\nu_{\hat \theta}^2
\right]\right\}\,.
\end{eqnarray}

The intrinsic curvature associated with the induced metric on $\Sigma$ can be easily calculated too. 
The nonvanishing components of the spatial Riemann tensor are given by
\begin{eqnarray}
\label{eq6.28}
R(\mathcal{N})_{r\phi r\phi}
  &=&g_{rr}g_{\phi\phi}[-E(n)_{\hat \theta\hat \theta}+E(n)_{\hat \phi\hat \phi}\nu_{\hat r}^2]
+K(\mathcal{N})_{11}K(\mathcal{N})_{33}+[K(\mathcal{N})_{13}]^2
\,, \nonumber\\
R(\mathcal{N})_{r\phi \theta\phi}
  &=&g_{\phi\phi}\sqrt{g_{rr}g_{\theta\theta}}[E(n)_{\hat r\hat \theta}+E(n)_{\hat \phi\hat \phi}\nu_{\hat r}\nu_{\hat \theta}]
-K(\mathcal{N})_{12}K(\mathcal{N})_{33}+K(\mathcal{N})_{13}K(\mathcal{N})_{23}
\,, \nonumber\\
R(\mathcal{N})_{r\theta r\phi}
  &=&g_{rr}\sqrt{g_{\theta\theta}g_{\phi\phi}}[(H(n)_{\hat r\hat r}+2H(n)_{\hat r\hat \theta}-\nu_nE(n)_{\hat r\hat \theta})\nu_{\hat r}
-(H(n)_{\hat r\hat \theta}-\nu_nE(n)_{\hat \theta\hat \theta})\nu_{\hat \theta}]\nonumber\\
&&
+K(\mathcal{N})_{11}K(\mathcal{N})_{23}+K(\mathcal{N})_{12}K(\mathcal{N})_{13}
\,,\nonumber\\
R(\mathcal{N})_{r\theta \theta\phi}
  &=&g_{\theta\theta}\sqrt{g_{rr}g_{\phi\phi}}[(-H(n)_{\hat r\hat \theta}+\nu_nE(n)_{\hat r\hat r})\nu_{\hat r}
+(H(n)_{\hat \theta\hat \theta}+2H(n)_{\hat r\hat r}+\nu_nE(n)_{\hat r\hat \theta})\nu_{\hat \theta}]\nonumber\\
&&
-K(\mathcal{N})_{12}K(\mathcal{N})_{23}-K(\mathcal{N})_{22}K(\mathcal{N})_{13}
\,,\nonumber\\
R(\mathcal{N})_{\theta\phi \theta\phi}
  &=&g_{\theta\theta}g_{\phi\phi}[-E(n)_{\hat r\hat r}+E(n)_{\hat \phi\hat \phi}\nu_{\hat \theta}^2]
+K(\mathcal{N})_{22}K(\mathcal{N})_{33}+[K(\mathcal{N})_{23}]^2
\,,\nonumber\\
R(\mathcal{N})_{r\theta r\theta}
  &=&g_{rr}g_{\theta\theta}\{
-E(n)_{\hat \phi\hat \phi}
+[E(n)_{\hat \theta\hat \theta}+\nu_n(2H(n)_{\hat r\hat \theta}-\nu_nE(n)_{\hat r\hat r})]\nu_{\hat r}^2\nonumber\\
&&
-2[(1+\nu_n^2)E(n)_{\hat r\hat \theta}+\nu_n(H(n)_{\hat r\hat r}-H(n)_{\hat \theta\hat \theta})]\nu_{\hat r}\nu_{\hat \theta}\nonumber\\
&&
+[E(n)_{\hat r\hat r}-\nu_n(2H(n)_{\hat r\hat \theta}+\nu_nE(n)_{\hat \theta\hat \theta})]\nu_{\hat \theta}^2
\}
\nonumber\\ 
&&
-K(\mathcal{N})_{11}K(\mathcal{N})_{22}+[K(\mathcal{N})_{12}]^2
\,,
\end{eqnarray}
where $E(n)$ and $H(n)$ denote the electric and magnetic parts of the spacetime Riemann tensor as measured by ZAMOs, respectively, defined by
\beq
E(n)_{\alpha\beta}=R_{\alpha\mu\beta\nu}n^\mu n^\nu\,,\qquad
H(n)_{\alpha\beta}=-[R^*]_{\alpha\mu\beta\nu}n^\mu n^\nu\,.
\eeq 
Their nonvanishing frame components are listed in Appendix A.
Note that conformally flat axisymmetric slices do not exist in general, as shown in Refs.~\cite{GP00,Kroon03}.

We know of no time slicings which take advantage of the extra freedom to depend on the polar angle $\theta$.
All of the interesting slicings known to us fall into the special case of ``spherical" slicings where $f=f(r)$ depends only on the Boyer-Lindquist radial coordinate $r$,
so that $\nu_{\hat\theta}=0$ and  ${\mathcal N}=\gamma(n+ \nu_{\hat r}\,\omega^{\hat r})$, i.e., it is just a boost of $n$ in the radial direction. Thus the new observers moving orthogonally to the new time slicing follow radial trajectories.
The induced metric on $\Sigma$ simplifies to
\beq
(\h_{ab})=\begin{pmatrix}
g_{rr}&0&0 \cr
0& g_{\theta\theta}& 0 \cr
0&0& g_{\phi\phi}\cr
\end{pmatrix}
+g_{t\phi}f_r
\begin{pmatrix}
0&0& 1\cr
0& 0& 0 \cr
1& 0& 0\cr
\end{pmatrix}
+g_{tt}f_r^2
\begin{pmatrix}
1&0&0 \cr
0&0& 0 \cr
0& 0&0\cr
\end{pmatrix}\,.
\eeq

The components of the 4-acceleration of observers having world lines orthogonal to the slices $\Sigma$ become
\begin{eqnarray}
a({\mathcal N})_1&=&\sqrt{g_{rr}}(\partial_{\hat r}\ln L)
=\sqrt{g_{rr}}\left[
a(n)_{\hat r}+\gamma^2\nu_{\hat r}\partial_{\hat r}\nu_{\hat r}
\right]\,, \nonumber\\
a({\mathcal N})_2&=&\sqrt{g_{\theta\theta}}(\partial_{\hat \theta}\ln L)
=\sqrt{g_{\theta\theta}}\left[
a(n)_{\hat \theta}+\gamma^2\nu_{\hat r}\partial_{\hat \theta}\nu_{\hat r}
\right]\,,
\end{eqnarray}
while the components of the extrinsic curvature tensor of each slice are
\begin{eqnarray}
K(\mathcal{N})_{11}&=& g_{rr}\gamma \left[ 
\nu_{\hat r}(W_{\hat r}-2\nu_n Z_{\hat r})+\partial_{\hat r}\nu_{\hat r}
\right]\,,\nonumber \\
K(\mathcal{N})_{12}&=& -\sqrt{g_{rr}g_{\theta\theta}}\gamma \nu_{\hat r}\left[ 
a(n)_{\hat \theta}+ \nu_n \theta(n)_{\hat \phi}{}_{\hat \theta}+ k(r,n)_{\hat \theta}
\right]\,,\nonumber \\
K(\mathcal{N})_{13}&=& \sqrt{g_{rr}g_{\phi\phi}}\gamma Z_{\hat r}\,,\nonumber\\
K(\mathcal{N})_{22}&=& g_{\theta\theta}\gamma \nu_{\hat r}k(\theta,n)_{\hat r}\,, \nonumber \\
K(\mathcal{N})_{23}&=& \sqrt{g_{\theta\theta}g_{\phi\phi}}\gamma \theta(n)_{\hat \phi}{}_{\hat \theta}\,,\nonumber\\
K(\mathcal{N})_{33}&=&g_{\phi\phi}\gamma\nu_{\hat r} k(\phi,n)_{\hat r}\,,
\end{eqnarray}
with
\beq
{\rm Tr}\,[K(\mathcal{N})]
=-\gamma\left[
\gamma^2\partial_{\hat r}\nu_{\hat r}+(a(n)_{\hat r}-k(\theta,n)_{\hat r}-k(\phi,n)_{\hat r})\nu_{\hat r}
\right]\,.
\eeq

Finally, the nonvanishing components of the spatial Riemann tensor (\ref{eq6.28}) simplify to
\begin{eqnarray}
R(\mathcal{N})_{r\phi r\phi}
&=&g_{rr}g_{\phi\phi}[-E(n)_{\hat \theta\hat \theta}+E(n)_{\hat \phi\hat \phi}\nu_{\hat r}^2]
+K(\mathcal{N})_{11}K(\mathcal{N})_{33}+[K(\mathcal{N})_{13}]^2
\,, \nonumber\\
R(\mathcal{N})_{r\phi \theta\phi}
&=&g_{\phi\phi}\sqrt{g_{rr}g_{\theta\theta}}\,E(n)_{\hat r\hat \theta}
-K(\mathcal{N})_{12}K(\mathcal{N})_{33}+K(\mathcal{N})_{13}K(\mathcal{N})_{23}
\,, \nonumber\\
R(\mathcal{N})_{r\theta r\phi}
&=&g_{rr}\sqrt{g_{\theta\theta}g_{\phi\phi}}\,[H(n)_{\hat r\hat r}+
2H(n)_{\hat r\hat \theta}-\nu_nE(n)_{\hat r\hat \theta}]\nu_{\hat r}
\nonumber\\
&&
+K(\mathcal{N})_{11}K(\mathcal{N})_{23}+K(\mathcal{N})_{12}K(\mathcal{N})_{13}
\,,\nonumber\\
R(\mathcal{N})_{r\theta \theta\phi}
&=&g_{\theta\theta}\sqrt{g_{rr}g_{\phi\phi}}\,[-H(n)_{\hat r\hat \theta}+\nu_nE(n)_{\hat r\hat r}]\nu_{\hat r}
\nonumber\\ 
&&
-K(\mathcal{N})_{12}K(\mathcal{N})_{23}-K(\mathcal{N})_{22}K(\mathcal{N})_{13}
\,,\nonumber\\
R(\mathcal{N})_{\theta\phi \theta\phi}
&=&-g_{\theta\theta}g_{\phi\phi}E(n)_{\hat r\hat r}
+K(\mathcal{N})_{22}K(\mathcal{N})_{33}+[K(\mathcal{N})_{23}]^2
\,,\nonumber\\
R(\mathcal{N})_{r\theta r\theta}
&=&g_{rr}g_{\theta\theta}\{
-E(n)_{\hat \phi\hat \phi}
+[E(n)_{\hat \theta\hat \theta}+\nu_n(2H(n)_{\hat r\hat \theta}-\nu_nE(n)_{\hat r\hat r})]\nu_{\hat r}^2\}\nonumber\\
&&
-K(\mathcal{N})_{11}K(\mathcal{N})_{22}+[K(\mathcal{N})_{12}]^2
\,.
\end{eqnarray}

Note that again as in the Schwarzschild case, geodesic slicings correspond to spatially constant lapse function $L=L_0$, and one could in principle consider the analogous hail, rain and drip coordinate systems.
However, in this case the simple geometrical properties of the intrinsic curvature are lost, and we will limit our discussion (see below) to the case of rain observers ($L_0=1$), which are identified with PG observers. 

\begin{figure}
\typeout{*** EPS figure 3}
\begin{center}
\includegraphics[scale=0.3]{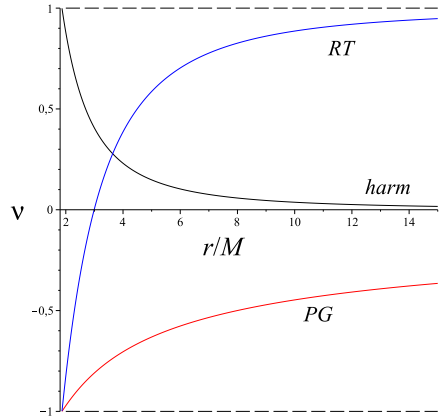}
\end{center}
\caption{Kerr spacetime slicings.
The behavior of the linear velocities associated with the special observer families considered in the text (i.e., harmonic, PG and RT) is shown for the choice of parameters $a/M=0.5$ and $\theta=\pi/2$, for which the outer horizon is located at $r_+\approx1.867M$.
}
\label{fig:3}
\end{figure}

Next some explicit examples are considered.
The behavior of the associated linear velocities is shown in Fig.~\ref{fig:3}.

% figure 3

\subsection{Harmonic slicings}

The harmonic condition (\ref{harm-cond}) implies
\beq
\partial_{\hat r}\nu_{\hat r}=[k(\theta,n)_{\hat r}+k(\phi,n)_{\hat r}]\nu_{\hat r}\,,
\eeq
so that 
\beq
\nu_{\hat r}=\frac{C\sin\theta}{\sqrt{g_{\theta\theta}g_{\phi\phi}}}\,,
\eeq
or equivalently
\beq
f_r=\frac{C}{\Delta}\,.
\eeq
Further integrating then gives
\beq
f(r)=\frac{C}{r_{+}-r_-}\ln\left|\frac{r-r_+}{r-r_{-}}\right|\,,
\eeq
Regularity of the new lapse function then requires $C=r_+^2+a^2$.

\begin{figure}
\typeout{*** EPS figure 4}
\begin{center}
\includegraphics[scale=0.3]{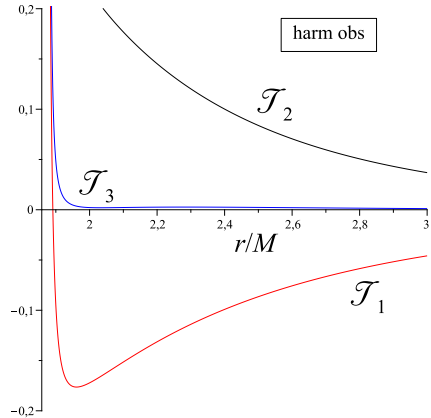}
\end{center}
\caption{Kerr spacetime, harmonic observers.
The traces (linear, quadratic and cubic) of the extrinsic curvature associated with harmonic observers are shown for the same choice of parameters as in Fig.~\ref{fig:3}. 
}
\label{fig:4}
\end{figure}

Harmonic slicings are neither intrinsically nor extrinsically flat. 
For example, the Ricci scalar has the following approximate expression
\beq
M^2\,R(h)=4 {\hat a}^2u^4[\cos^2\theta+(3\cos^2\theta - 1)u]+O(u^6)\,.
\eeq
The traces (linear, quadratic, cubic) of the extrinsic curvature associated with harmonic observers are shown in Fig.~\ref{fig:4}.

% figure 4

\subsection{Geodesic slicings}

Geodesic slicings are characterized by vanishing acceleration, i.e., by constant lapse function $L=L_0$ of the spacetime metric written in adapted coordinates.
Therefore, they are identified by the condition 
\beq
\gamma=\frac{L_0}{N}\,,\qquad
\nu_{\hat r}=\pm\sqrt{L_0^2-N^2}\,,
\eeq
the same condition as in the corresponding Schwarzschild spacetime. 

When $L_0=1$ these slicings are associated with the PG observers, whose world lines are a timelike congruence with both vanishing acceleration and vorticity in the Kerr spacetime as well \cite{Painleve21,Gullstrand22,Doran00,natario,herrero,geosli}.
Their relative velocity with respect to the ZAMOs is
\beq
\nu_{\hat r}=-\sqrt{1-N^2}=-\sqrt{ \frac{2Mr(r^2+a^2)}{\Sigma g_{\phi\phi}}}\sin \theta\,,
\eeq
corresponding to radially infalling observers,
leading to 
\beq
f(r)=-\int\frac{\sqrt{2Mr(r^2+a^2)}}{\Delta}\,\rmd r\,.
\eeq

In contrast to the Schwarzschild case, the geodesic condition does not imply intrinsic flatness since
the Ricci tensor of the induced metric does not vanish identically.
It can be expressed in terms of the extrinsic curvature tensor as in Eq.~(\ref{geosli}).
The spatial curvature scalar is given by
\beq
R(\h)=\frac{2a^2Mr(3\cos^2\theta -1)}{\Sigma^3}\,.
\eeq
Thus the geometry associated with the PG observers in the Kerr spacetime is neither intrinsically nor extrinsically flat.

\begin{figure}
\typeout{*** EPS figure 5}
\begin{center}
\includegraphics[scale=0.3]{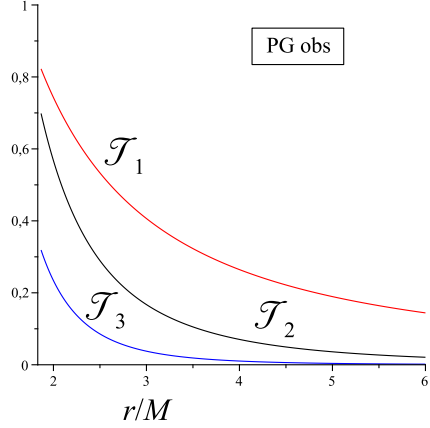}
\end{center}
\caption{Kerr spacetime, PG observers.
The rescaled traces (linear, quadratic and cubic) of the extrinsic curvature associated with the PG observers are shown for the same choice of parameters as in Fig.~\ref{fig:3}. 
}
\label{fig:5}
\end{figure}

For completeness we list also the linear, quadratic and cubic invariants of $K({\mathcal N}_{\rm PG})$:
\begin{eqnarray}
M\, {\rm Tr}\,[K({\mathcal N}_{\rm PG})]&=&
\frac{u^{3/2}}{\tilde \Sigma[2(1+\hat a^2u^2)]^{1/2}}
(3+{\hat a}^2u^2)
\,,\nonumber \\
M^2\, {\rm Tr}\,[K({\mathcal N}_{\rm PG})^2]&=&
\frac{u^3}{2{\tilde \Sigma}^3(1+\hat a^2u^2)}
[9+(10-3\cos^2\theta){\hat a}^2u^2\nonumber\\
&&+(5-6\cos^2\theta){\hat a}^4u^4+\cos^2\theta\,{\hat a}^6u^6]
\,,\nonumber \\
M^3\, {\rm Tr}\,[K({\mathcal N}_{\rm PG})^3]&=&
\frac{u^{9/2}}{{\tilde \Sigma}^4[2(1+\hat a^2u^2)]^{3/2}}
[15+3(7-\cos^2\theta){\hat a}^2u^2\nonumber\\
&&+21\sin^2\theta{\hat a}^4u^4
+(7-9\cos^2\theta){\hat a}^6u^6+\cos^2\theta\,{\hat a}^8u^8]
\,.
\end{eqnarray}
The eigenvalues of $K({\mathcal N}_{\rm PG})$ are also easily evaluated
\begin{eqnarray}
M\lambda_0 &=&
\frac{\sqrt{2}u^{3/2}}{\tilde \Sigma\sqrt{1+\hat a^2u^2}}
\,,  \\
M\lambda_{\pm} &=&
\frac{M\lambda_0}{4}(1+\hat a^2u^2)
\pm\left(\frac{u}{2{\tilde \Sigma}}\right)^{3/2}[9+3(3-5\cos^2\theta){\hat a}^2u^2+\cos^2\theta\,{\hat a}^4u^4]^{1/2}
\,. \nonumber
\end{eqnarray}
Note that $\lambda_0$ is always positive.
In the Schwarzschild limit the above expressions reduce to $M\lambda_0=M\lambda_+=-2M\lambda_-=\sqrt{2}u^{3/2}$.
The traces (linear, quadratic, cubic) of the extrinsic curvature associated with PG observers are shown in Fig.~\ref{fig:5}.

% figure 5

\subsection{Hyperboloidal slicings}

Numerical relativity computations (like those concerning outgoing gravitational radiation) use preferred hyperboloidal spacetime slicings. 
A general framework for the construction of hyperboloidal coordinates with scri-fixing (i.e., with an explicit prescription to fix the coordinate location of null infinity) for stationary, weakly asymptotically flat spacetimes (including black hole spacetimes) was first developed by Moncrief \cite{moncrief} (see also  \cite{Zenginoglu:2007jw,Zenginoglu:2008uc}). 

Recently this method has been successfully applied to the numerical investigation of tail decay rates in a Kerr spacetime by Racz and Toth  (RT) \cite{Racz:2011qu} (see also Refs.~\cite{Bernuzzi:2011aj,Yang:2013uba,Harms:2014dqa}).
Their construction starts from the standard Boyer-Lindquist coordinates $(t,r,\theta,\phi)$, passing then to ingoing Kerr coordinates $(\bar t, r, \theta ,\bar \phi)$
such that
\begin{eqnarray}
\bar t = t-r +\int \frac{r^2+a^2}{\Delta}\, \rmd r\,,\qquad 
\bar \phi =\phi + \int \frac{a}{\Delta}\,\rmd r\,.
\end{eqnarray}
Finally, the time and radial coordinates $\bar t$ and $r$ are replaced by the new time coordinate $T$ and the compactified radial coordinate $R$, implicitly defined by 
\begin{eqnarray}
\bar t = T-4M \left[\ln \left(1-\frac{R^2}{M^2}\right)-\frac14\frac{M^2+R^2}{M^2-R^2}\right]\,,\qquad 
r =\frac{2M^2 R}{M^2-R^2}\,.
\end{eqnarray}
The advantage of using RT coordinates $(T,R,\theta,\bar \phi)$ is that the time slices are horizon penetrating and connect to future null infinity, so that no boundary conditions are needed.
The relation for the radial coordinate $r=r(R)$ can be easily inverted, leading to
\beq
\label{Rdef}
\frac{R}{M}=\sqrt{1+\frac{M^2}{r^2}}-\frac{M}{r}
\,.
\eeq
Solving for $T$ then gives $T=t-f(r)$ with
\beq
f(r)=r-\int  \frac{r^2+a^2}{\Delta}\, \rmd r -4M\left[\ln \left(1-\frac{R^2}{M^2}  \right)-\frac14\frac{M^2+R^2}{M^2-R^2}\right]\,.
\eeq
RT observers thus form an irrotational congruence ${\mathcal N}_{\rm RT}$ of world lines moving radially with respect to ZAMOs, with a relative velocity given by
\beq
\nu_{\hat r}=\frac{N}{\sqrt{g_{rr}}}\left[
-\frac{2Mr}{\Delta}+\frac{4M}{r}\left(
1+\frac{\frac14-\frac{M^2}{r^2}}{\frac{R}{r}+\frac{M^2}{r^2}}\right)
\right]\,,
\eeq
with $R=R(r)$ as defined in Eq.~(\ref{Rdef}).
It is easy to show also that 
\beq
\lim_{r\to r_+}\nu_{\hat r}=-1\,,\quad 
\lim_{r\to \infty}\nu_{\hat r}=1\,,
\eeq
with $\nu_{\hat r}$ passing monotonically from $-1$ to $+1$, irrespective of the value of $\theta$ (see Fig. \ref{fig:3}).

\begin{figure}
\typeout{*** EPS figure 6}
\begin{center}
\includegraphics[scale=0.3]{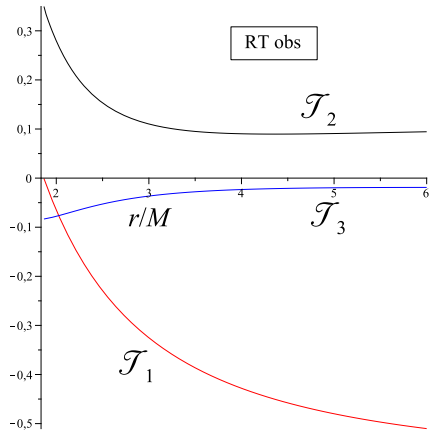}
\end{center}
\caption{Kerr spacetime, RT observers.
The traces (linear, quadratic and cubic) of the extrinsic curvature associated with RT observers are shown for the same choice of parameters as in Fig.~\ref{fig:3}. 
}
\label{fig:6}
\end{figure}

The traces (linear, quadratic and cubic) of the extrinsic curvature are shown in Fig.~\ref{fig:6}.

% figure 6

\section{Concluding remarks}

Special congruences of timelike world lines in Kerr spacetime have been studied extensively in the literature (see, e.g., Refs.~\cite{idcf1,idcf2} and references therein), while 
in contrast there exist only a few interesting spacetime spacelike slicings that have been investigated.
The most natural one is the maximal spacelike slicing associated with the Boyer-Lindquist time coordinate, whose properties are best described in terms of the orthogonal congruence of timelike world lines threading this slicing associated with the so-called ZAMO observers.
This congruence is accelerated and shearing but (locally) nonrotating, but the induced metric on the slices does not have any special properties.
Nevertheless, other interesting slicings can be found compatible with the  Killing symmetries of the spacetime, motivated either by geometrical or analytical considerations.

We have developed here a general framework for analyzing all possible stationary slicings allowed in a given stationary geometry, then specialized our results to black hole spacetimes, reviewing the various familiar examples which can be found in the literature.
Geometrical properties of these slicings can be characterized either in terms of those of the kinematical properties of their orthogonal timelike congruences interpreted as the world lines of a family of stationary observers, or in terms of intrinsic or extrinsic curvature properties of the slicing itself. 
We have considered in detail the case of slicings associated with a new time coordinate function of the Boyer-Lindquist coordinates $(t,r)$. The case of geodesic observers includes the so-called Painlev\'e-Gullstrand slicing of Kerr. 
The latter are even more special in the limiting Schwarzschild case where the slices are also intrinsically flat. Finally
we have also presented examples of ``analytical'' slicings of black hole spacetimes, like those associated with harmonic and hyperboloidal time gauges.

\appendix

\section{ZAMO relevant quantities in the Kerr spacetime}
\label{appzamos}

We list below the non-vanishing components of the electric and magnetic parts of the Riemann tensor as well as the relevant kinematical quantities as measured by ZAMOs.

The components of the acceleration and expansion vectors are given by
\begin{eqnarray}
a(n)^{\hat r}&=&\frac{M\sin^2\theta}{\Sigma^{5/2}\sqrt{\Delta}g_{\phi\phi}}\left\{-2r^2 (r^2+a^2)\Delta +\Sigma[(r^2+a^2)^2-4Mr^3]\right\}\,,\nonumber\\
a(n)^{\hat \theta}&=&-\frac{2a^2Mr}{\Sigma^{5/2}g_{\phi\phi}}(r^2+a^2) \sin^3\theta \cos\theta\,,\nonumber\\
\theta(n)_{\hat r\hat \phi}&=&-\frac{Ma\sin^3\theta}{\Sigma^{5/2}g_{\phi\phi}}[2r^2(r^2+a^2)+\Sigma(r^2-a^2)]
\,,\nonumber\\
\theta(n)_{\hat \theta\hat \phi}&=&\frac{2a^3Mr}{\Sigma^{5/2}g_{\phi\phi}} \sqrt{\Delta}\cos\theta\sin^4\theta
\,,
\end{eqnarray}
and the curvature components are
\begin{eqnarray}
k(r,n)^{\hat r}&=&\frac{\Sigma (r-M)-r\Delta}{\Sigma^{3/2}\sqrt{\Delta}}\,,\qquad
k(r,n)^{\hat \theta}=\frac{a^2 \sin\theta \cos\theta}{\Sigma^{3/2}}\,,\nonumber\\
k(\theta,n)^{\hat r}&=&-\frac{r\sqrt{\Delta}}{\Sigma^{3/2}}\,,\qquad
k(\theta,n)^{\hat \theta}=\frac{a^2 \sin\theta \cos\theta}{\Sigma^{3/2}}\,,\nonumber\\
k(\phi,n)^{\hat r}&=&\frac{\sqrt{\Delta}\sin^2\theta}{\Sigma^{5/2}g_{\phi\phi}}[-\Sigma^2 (r-M)-M\Sigma (3r^2+a^2)+2Mr^2(r^2+a^2)]
\,,\nonumber\\
k(\phi,n)^{\hat \theta}&=&-\frac{\cos\theta\sin\theta}{\Sigma^{5/2}g_{\phi\phi}}[\Sigma^2\Delta +2Mr(r^2+a^2)^2]
\,.
\end{eqnarray}
Finally, the nontrivial components of the electric and magnetic parts of the Riemann tensor with respect to ZAMOs are given by
\begin{eqnarray}
\label{E_H}
E_{\hat r \hat r}&=&\partial_{\hat r}a(n)_{\hat r}+a(n)_{\hat r}^2-3\theta(n)_{\hat r\hat \phi}^2-a(n)_{\hat \theta}k(r,n)^{\hat \theta}
\,,\nonumber\\
E_{\hat \theta \hat \theta}&=&-E_{\hat r \hat r}+\partial_{\hat r}k(\theta,n)^{\hat r}-[k(\theta,n)^{\hat r}]^2
+\partial_{\hat \theta}k(r,n)^{\hat \theta}-[k(r,n)^{\hat \theta}]^2
\,,\nonumber\\
E_{\hat r \hat \theta}&=&\partial_{\hat \theta}a(n)_{\hat r}+a(n)_{\hat \theta}[a(n)_{\hat r}+k(\theta,n)^{\hat r}]-3\theta(n)_{\hat r\hat \phi}\theta(n)_{\hat \theta\hat \phi}
\,,\nonumber\\
H_{\hat r \hat r}&=&\partial_{\hat \theta}\theta(n)_{\hat r\hat \phi}-\theta(n)_{\hat r\hat \phi}k(\phi,n)^{\hat \theta}
+\theta(n)_{\hat \theta\hat \phi}[k(\theta,n)^{\hat r}-k(\phi,n)^{\hat r}]
\,,\nonumber\\
H_{\hat r \hat \theta}&=&-\partial_{\hat r}\theta(n)_{\hat r\hat \phi}+2\theta(n)_{\hat r\hat \phi}k(\phi,n)^{\hat r}
+\theta(n)_{\hat \theta\hat \phi}k(r,n)^{\hat \theta}
\,,\nonumber\\
H_{\hat \phi \hat \phi}&=&\theta(n)_{\hat r\hat \phi}[a(n)_{\hat \theta}+k(\phi,n)^{\hat \theta}]
-\theta(n)_{\hat \theta\hat \phi}[a(n)_{\hat r}+k(\phi,n)^{\hat r}]
\,,
\end{eqnarray}
in terms of the above kinematical quantities.

In the equatorial plane, the acceleration and expansion and curvature vectors are all directed along the radial direction, with components
\beq
a(n)^{\hat r}=\frac{M}{r^2\sqrt{\Delta}}\frac{(r^2+a^2)^2-4a^2Mr}{r^3+a^2r+2a^2M}\,,\quad
\theta_\phi(n)^{\hat r}=-\frac{aM(3r^2+a^2)}{r^2(r^3+a^2r+2a^2M)}\,,
\eeq
and
\begin{eqnarray}
\kappa(r,n)^{\hat r}&=&\frac{Mr-a^2}{r^2\sqrt{\Delta}}\,,\qquad
\kappa(\theta,n)^{\hat r}=-\frac{\sqrt{\Delta}}{r^2}\,,\nonumber\\
k(\phi,n)^{\hat r}&=&-\frac{(r^3-a^2M)\sqrt{\Delta}}{r^2(r^3+a^2r+2a^2M)}\,,
\end{eqnarray}
respectively, whereas the nonvanishing frame components (\ref{E_H}) of the electric and magnetic parts of the Riemann tensor reduce to
\begin{eqnarray}
E_{\hat r \hat r}&=& -\frac{M(2 r^4+5 r^2 a^2-2 a^2 M r+3 a^4)}{r^4 (r^3+a^2 r+2 a^2 M)}\,, \quad
E_{\hat \theta \hat \theta}=-E_{\hat \phi \hat \phi}- E_{\hat r \hat r}\,,\quad
E_{\hat \phi \hat \phi}=\frac{M}{r^3}\,,\nonumber\\
H_{\hat r \hat \theta}&=&  -\frac{3 M a (r^2+a^2) \sqrt{\Delta}}{r^4 (r^3+a^2 r+2 a^2 M)}\,.
\end{eqnarray}

\section*{Acknowledgements}
DB thanks Dr. A. Nagar and E. Harms for informative discussions concerning the RT coordinate system in Kerr.
EB is financially supported by the CAPES-ICRANet program (BEX 13956/13-2).

\end{document}